\documentclass[lettersize,journal]{IEEEtran}
\usepackage{amsmath,amsfonts,amssymb}
\usepackage{algorithmic}
\usepackage{algorithm}
\usepackage{array}
\usepackage[caption=false,font=normalsize,labelfont=sf,textfont=sf]{subfig}
\usepackage{textcomp}
\usepackage{stfloats}
\usepackage{url}
\usepackage{verbatim}
\usepackage{graphicx}
\usepackage{epstopdf}
\usepackage{cite}
\usepackage[english]{babel}
\usepackage{amsthm}
\usepackage{array}
\usepackage[caption=false,font=normalsize,
labelfont=sf,textfont=sf]{subfig}
\usepackage{textcomp}
\usepackage{stfloats}
\usepackage{url}
\usepackage{balance}
\usepackage{multirow}
\usepackage{xcolor}

\newtheorem{theorem}{Theorem}
\newtheorem{lemma}{Lemma}
\newtheorem{proposition}{Proposition}
\newtheorem{definition}{Definition}
\newtheorem{property}{Property}
\newtheorem{assumption}{Assumption}
\newcommand\Item[1][]{%
  \ifx\relax#1\relax  \item \else \item[#1] \fi
  \abovedisplayskip=0pt\abovedisplayshortskip=0pt~\vspace*{-\baselineskip}}

\begin{document}

\title{Downlink Channel Covariance Matrix Reconstruction for FDD Massive MIMO Systems with Limited Feedback}

\author{Kai Li, Ying Li, Lei Cheng, Qingjiang Shi, and Zhi-Quan Luo, \IEEEmembership{Fellow, IEEE}
\thanks{Kai Li, Ying Li and Zhi-Quan Luo are with the Chinese
University of Hong Kong, Shenzhen 518172, China, and also with Shenzhen Research Institute of Big Data, Shenzhen 518172,
China (e-mail: kaili4@link.cuhk.edu.cn; yingli5@link.cuhk.edu.cn; luozq@cuhk.edu.cn).}
\thanks{Lei Cheng is with the College of Information Science and Electronic Engineering at Zhejiang University, Hangzhou 310058, China. He is also with Shenzhen Research Institute of Big Data, Shenzhen 518172, China (e-mail: lei\_cheng@zju.edu.cn).}
\thanks{Qingjiang Shi is with the School of Software Engineering at Tongji University, Shanghai 201804, China, and also with Pazhou Lab, Guangzhou 510310, China (e-mail: shiqj@tongji.edu.cn).}
}

\markboth{Journal of \LaTeX\ Class Files,~Vol.~14, No.~8, August~2021}%
{Shell \MakeLowercase{\textit{et al.}}: A Sample Article Using IEEEtran.cls for IEEE Journals}

\IEEEpubid{0000--0000/00\$00.00~\copyright~2021 IEEE}

\maketitle

\begin{abstract}
The downlink channel covariance matrix (CCM) acquisition is the key step for the practical performance of massive multiple-input and multiple-output (MIMO) systems, including beamforming, channel tracking, and user scheduling. However, this task is challenging in the popular frequency division duplex massive MIMO systems with Type I codebook due to the limited channel information feedback. In this paper, we propose a novel formulation that leverages the structure of the codebook and feedback values for an accurate estimation of the downlink CCM. Then, we design a cutting plane algorithm to consecutively shrink the feasible set containing the downlink CCM, enabled by the careful design of pilot weighting matrices. Theoretical analysis shows that as the number of communication rounds increases, the proposed cutting plane algorithm can recover the ground-truth CCM. Numerical results are presented to demonstrate the superior performance of the proposed algorithm over the existing benchmark in CCM reconstruction.  
\end{abstract}


\section{Introduction}

\subsection{Background and Challenge}
\label{sec:background}
\IEEEPARstart{I}{n} the fifth generation (5G) cellular systems and beyond (e.g., 5.5G and 6G \cite{5_5G}), massive multiple-input and multiple-output (MIMO) has become a key enabling technology \cite{massive_MIMO}. By harnessing a large number of antennas at the base station (BS), the efficient use of spectral resources \cite{MIMO_1} and the mitigation of inter-cell interference \cite{MIMO_2} can be performed via simple algorithms, thus facilitating the implementations on hardware. Other prominent functionalities of massive MIMO include joint spatial division and multiplexing \cite{MIMO_3}, optimal user scheduling \cite{MIMO_4}, channel tracking \cite{MIMO_5}, and so forth. 

To realize the potential merits of massive MIMO for aforementioned tasks \cite{MIMO_1,MIMO_2,MIMO_3,MIMO_4,MIMO_5}, downlink channel covariance matrix (CCM) is often required at BS as indispensable prior knowledge. It measures how downlink channels are correlated across different antennas, and varies quite slowly compared to the instantaneous channel realizations \cite{channel_est_1}. Therefore, downlink CCM is essential for the design of long-term statistically  adaptive algorithms in practical wireless systems \cite{5_5G}, and thus demands accurate acquisition. 

Downlink CCM acquisition in frequency-division duplex (FDD) wireless systems is much more challenging than the counterpart in time-division duplex (TDD) systems, due to the lack of channel reciprocity property \cite{FDD}. \textcolor{black}{Despite the challenge, FDD shows advantages over TDD in several aspects, including data rates, cellular coverage, network investment, and time synchronization.} To tackle the difficulty caused by lacking channel reciprocity in a communication-efficient manner, existing real-life FDD 5G systems adopt codebook based limited feedback schemes \cite{feedback_scheme}. Their performances are encouraging, as evidenced by recent research works \cite{FDD_feedback_scheme_1,FDD_feedback_scheme_3} and industry report \cite{FDD_feedback_scheme_4}, thus forming another major path enabling the promising functionalities of massive MIMO, parallel to the TDD-based one. In the codebook based limited feedback schemes, the BS and the user equipment (UE) share a judiciously designed codebook (e.g., Type I codebook in the 3GPP standard \cite{code1}), which is a set of vectors (a.k.a. codewords) that approximate instantaneous channels. At the UE, after acquiring the CCM via the downlink training process, it utilizes the codebook to ``encode'' CCM into a few scalars, and then feeds these scalars, rather than the whole CCM, back to the BS. Although with light overhead, this scheme results in a challenging task at the BS: \emph{how to reconstruct the CCM, possibly with a large number of entries, from only a few feedback values?}
 \vspace{-0.1cm}
\subsection{Related Works}
\label{sec:related_works}
 \IEEEpubidadjcol
 \textcolor{black}{Based on the assumption that the directional information tends to be reciprocal, many works, such as \cite{ channel_reciprocity_based_3,channel_reciprocity_based_4,channel_reciprocity_based_5}, estimate the downlink CCM using the uplink channel directional (angular) information. In particular, this type of approach poses a critical challenge in the channel parameter acquisition of uplink channels, especially for the channel consisting of many paths (e.g., CDL channel model \cite[Sec. 7.7.1]{R1}). Therefore, previous works \cite{ channel_reciprocity_based_3,channel_reciprocity_based_4,channel_reciprocity_based_5} usually consider the communication system with the sparse channel, based on which they exploit the advances in array signal processing to tackle the challenge of estimating channel parameters. However, the real-life 5G channel is usually non-sparse \cite[Sec. 7.7.1]{R1}, and thus much effort has to be put into uplink channel parameter estimation.}

On the other hand, several previous works investigated downlink CCM estimation at the UE using pilot signals \cite{channel_est_UE_1, channel_est_UE_2, channel_est_UE_3}. For example, in \cite{channel_est_UE_1}, based on the orthogonal assumption of channels, it was shown that each channel vector can be characterized by an eigenvector of the sample covariance matrix computed by the received data at the UE. Then, with the aid of pilot signals, channel vectors and their corresponding downlink CCM can be acquired via eigen-vector decomposition (EVD). However, since the pilot signals are contaminated by the inter-cell inference, the channel vectors are only approximately orthogonal in practice. To mitigate pilot contamination, research work \cite{channel_est_UE_2} proposed a Bayesian channel estimation method assuming that the BS coordination among cells is viable. In addition, after taking the  transceiver's hardware impairment into account, recent work \cite{channel_est_UE_3} proposed a robust CCM estimation algorithm. 

The algorithms mentioned above are performed {\it at the UE} using the received data and pilot signals, and thus {\it cannot} address the CCM reconstruction challenge {\it at the BS}, in which only a few feedback values are available. Several early attempts have been made for different limited feedback schemes \cite{feedback_1,feedback_2,nikos}. For instance, in \cite{feedback_1}, the received pilot signal at the UE is fed back to the BS. By assuming the sparsity of massive MIMO channels, a compressed sensing based method was proposed to estimate downlink CCM. Adopting the same feedback scheme as \cite{feedback_1}, the work \cite{feedback_2} proposed a two-stage weighted block $l_1$ minimization based CCM reconstruction algorithm. In \cite{nikos}, the signal-to-noise ratio (SNR) measure is fed back to the BS, based on which a cutting plane method was utilized for CCM reconstruction and subsequent beamforming. More recently, the deep learning model was employed to realize  CCM compression at the UE and CCM reconstruction at the BS \cite{dl_based_5G}. For various limited feedback schemes mentioned above, they are not as widely adopted as the codebook based scheme (introduced in Section \ref{sec:background}) in real-world 5G systems, see, e.g., the 3GPP standard \cite{code1}.  

Using the codebook based feedback mechanism, there are several works on downlink CCM reconstruction \cite{quantization1,quantization2,patent}, which performances heavily rely on the selected codebook.\textcolor{black}{In \cite{quantization1,quantization2}, codebooks are designed via quantization techniques, which involve mapping a continuous set of values to a discrete set of values\cite{gray1984vector}. This technique is widely applied to design codebooks that enable the efficient signal representation and transmission\cite{choi2013noncoherent,love2003grassmannian}.} However,  codebook is not frequently altered in practical 5G systems. Therefore, other works mainly focus on the exploitation of codebook. For instance, the most closely related work \cite{patent} approximated the CCM via the codewords indicated by feedback values, which can be viewed as an approximation of the principal eigenvectors of the downlink CCM. So far, given a pre-defined codebook and limited feedback values, the downlink CCM reconstruction problem has not been well formulated nor solved in a principled fashion, especially in view of the 3GPP standard \cite{code1}. 
  \vspace{-0.1cm}
\subsection{Contributions}
To fill in this gap, we propose a novel algorithm that reconstructs the downlink CCM for Type I codebook (in the 3GPP standard \cite{code1}) based limited feedback FDD massive MIMO systems. The major contributions of this paper are summarized as follows.

\noindent $\bullet$ \textbf{Principled Problem Formulation.} To the best knowledge of the authors, it is the {\it first} time that the downlink CCM reconstruction problem in the context of Type I codebook based FDD wireless systems has been formulated in a principled way, by leveraging the structures of codebook and feedback values. The resulting problem aims at consecutively squeezing the size of the feasible set, which is characterized by equality/inequality constraints. 

\noindent $\bullet$ \textbf{Effective Algorithm Design.} Due to numerous constraints, evaluating the size of the feasible set is challenging (and even intractable). To get over this hurdle, this paper proposes an effective algorithm to optimize pilot weighting matrices such that the feasible set can be consecutively reduced. It is the {\it first} time that these pilot weighting matrices have been judiciously optimized for accurate CCM reconstruction. Numerical experiments based on channel samples from QUAsi Deterministic RadIo channel GenerAtor (QuaDRIGa)\footnote{https://quadriga-channel-model.de.} have confirmed the excellent performance of the proposed algorithm. 

\noindent $\bullet$ \textbf{Convergence Analysis.} We provide the convergence characterization of the proposed algorithm. It is shown that the algorithm can exactly recover the ground-truth CCM if the communication round goes to infinity. This indicates that in practice, given adequate feedback values, the proposed algorithm can reconstruct CCM with a relatively high accuracy. 

Part of the work has appeared in IEEE ICASSP 2021 \cite{ICASSP}. The current paper contains convergence analysis and more extensive experiment results that were missing in the conference version \cite{ICASSP}. \textcolor{black}{Furthermore, the codes of the proposed algorithm and baseline can be found in GitHub (see link: https://github.com/wamcs/CCM-Reconstruction).}
 \vspace{-0.1cm}
\subsection{The Structure of This Paper and Notations}
The remainder of this paper is organized as follows. In Section \ref{sec:system}, the system model and problem formulation are introduced, based on which the CCM reconstruction algorithm is developed and analyzed in Section \ref{sec:channel_reconstruction}. 
In Section \ref{sec:numerical}, numerical results using channel samples from QuaDRIGa are presented to show the excellent performance of the proposed algorithm. Finally, conclusions are drawn in Section
\ref{sec:conclusion}. 

Throughout the paper, we use boldface uppercase letters to denote matrices and boldface lowercase letters to denote column vectors. The superscript $(\cdot)^H$ is adopted to denote the Hermitian (conjugate) transpose matrix operator. The set containing $N$ elements is denoted by $\{(\cdot)_n\}_{n=1}^N$, where $n$ is the index of the element. The intersection of $N$ sets is denoted by $\bigcap_{n=1}^N \{\cdot \}_n$, where $n$ is the index of the set. The symbol $\subseteq$ denotes the subset relationship. In addition, $\mathrm{Tr}(\cdot)$, $\mathrm{rank}(\cdot)$, $\mathbb{E}[\cdot]$, and $\mathrm{diag}(\cdot)$ denote the trace, rank, expectation, and diagonalization operators respectively. Particularly, $\langle\cdot,\cdot\rangle_F$ denotes matrix inner product operator, which is equivalent to the trace of the product of two matrices, that is, $\langle \mathbf{A},\mathbf{B}\rangle_F = \mathrm{Tr}(\mathbf{A}\mathbf{B})$.

\section{System Model And Problem Formulation}
\label{sec:system}
\begin{figure}[!t]
    \centering
    \includegraphics[width= 3.5 in]{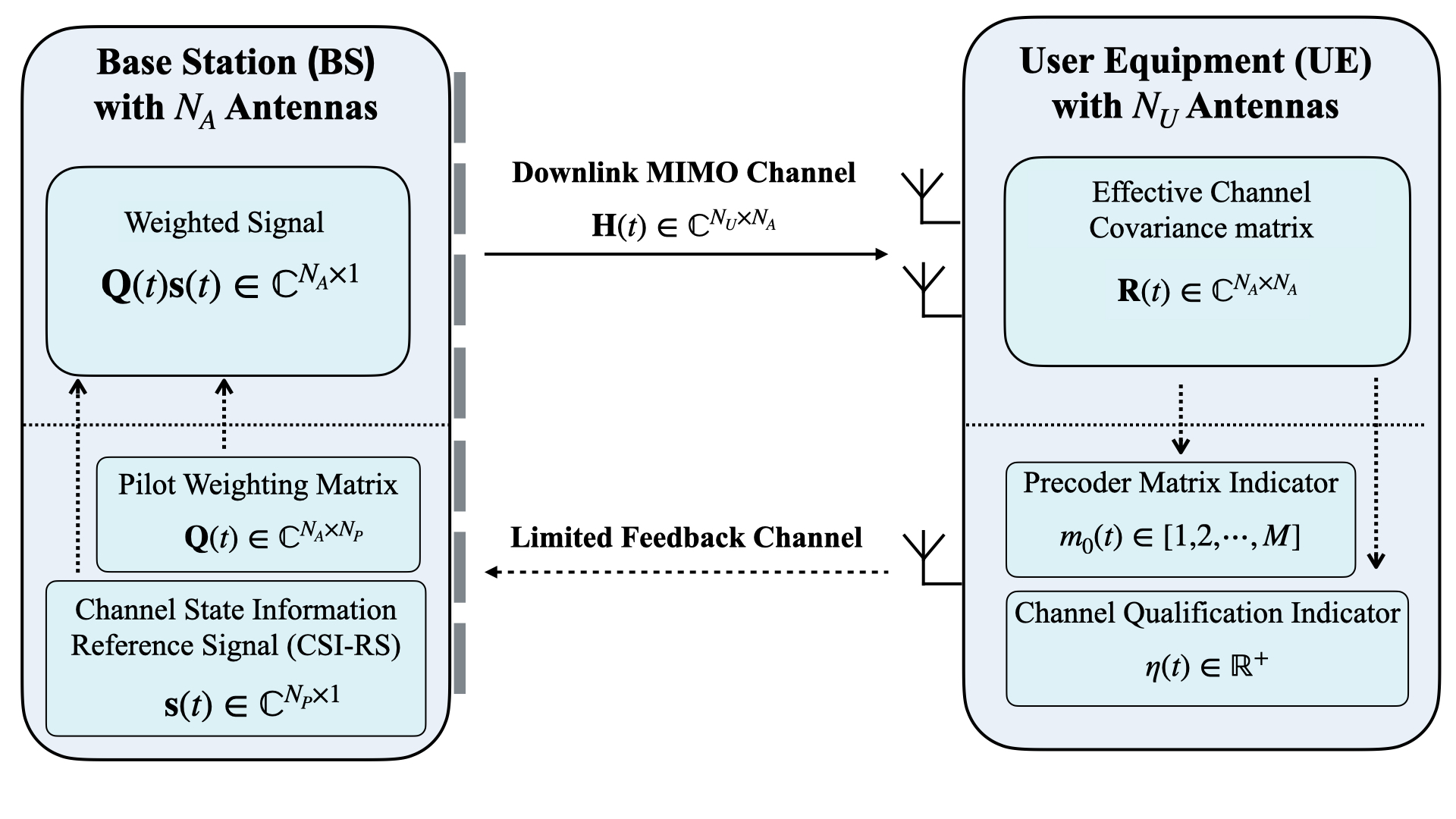}
    \caption{Block diagram of the downlink CCM reconstruction via feeding back precoding matrix indicator and channel quality indicator values.}
    \label{fig_system}
    \vspace{-0.3cm}
\end{figure}

As shown in Fig. \ref{fig_system},  a point-to-point MIMO link consists of a BS with $N_A$ antennas and a UE with $N_U$ antennas. In the $t$-th communication round, the downlink channel matrix between the BS and the UE is denoted by $\mathbf{H}(t) \in \mathbb C^{N_U \times N_A}$. Moreover, within $T$ consecutive communication rounds, the downlink CCM $\mathbf{C} = \mathbb{E}[\mathbf{H}(t)^H \mathbf{H}(t)] \in \mathbb C^{N_A \times N_A}$ is assumed to remain unchanged.

Following the 5G new radio (NR) standard \cite{code1}, the channel state information reference signal (CSI-RS), denoted by $\mathbf s(t) \in \mathbb C^{N_P \times 1}$, is adopted to assist the UE to acquire the CSI in the $t$-th communication round. On the other hand,  since the number of CSI-RS ports $N_P$ might be smaller than the number of antennas $N_A$ at the BS (i.e., $N_P \leq N_A$), the CSI-RS $\mathbf s(t)$ needs to be reshaped to match the dimension of antennas at the BS. Particularly, by using the pilot weighting matrix $\mathbf Q(t) \in \mathbb C^{N_A \times N_P}$ (also known as virtual antenna matrix), BS transmits the signal $\mathbf Q(t) \mathbf s(t) \in \mathbb{C}^{N_A \times 1}$ to the UE through channel $\mathbf{H}(t)$. At the UE, the received data takes the following form: 
\begin{align}
    \mathbf y(t) = \mathbf{H}(t) \mathbf Q(t) \mathbf s(t) + \mathbf z(t), ~~ t = 1, \cdots, T, \label{received_data}
\end{align} 
where $\mathbf z(t) \sim \mathcal {CN}(\mathbf z(t) | \mathbf 0, \sigma^2 \mathbf I_{N_U})$ is the additive white Gaussian noise (AWGN). \textcolor{black}{Using CSI-RS $\mathbf s(t)$ and received signal $ \mathbf y(t)$, the UE can estimate the effective channel, modeled as 
\begin{align}
  \mathbf{H}_e(t)=\mathbf{H}(t) \mathbf{Q}(t)+\mathbf{E}(t),\quad t= 1,\cdots,T, \label{effective_channel}
\end{align}
where $\mathbf{E}_{ij}(t) \sim \mathcal{N}(0,\sigma_{e_t})$ is the estimation error. In current communication systems, the estimated effective channel has high accuracy. For simplicity, we first ignore the error term in the algorithm development, and then investigate its effect in Section \ref{sec:numerical}. Furthermore, since $\mathbf{H}_e(t)$ is assumed to be constant during the coherent interval, the receiver-side effective CCM is computed by
\begin{align}
   \mathbf R(t) \triangleq  \mathbf H_e (t)^H \mathbf H_e (t)  &= \mathbf Q(t)^H \mathbf{C} \mathbf Q(t),  t = 1, \cdots, \!T \label{effective_channel_covariance_matrix}
\end{align}
via various algorithms\cite{channel_est_UE_1, channel_est_UE_2, channel_est_UE_3}\footnote{\textcolor{black}{It is worth noticing that the 3GPP protocol \cite{R3} states that the user needs to measure downlink CSI via received CSI-RS, but the protocol does not specify how to measure CSI, thus attracting wide research attention.}}(see Section \ref{sec:related_works}}).

At the UE, instead of directly transmitting the CCM $\mathbf R(t)$ back to the BS via the reverse link, Type I codebook $\mathcal V \triangleq \{\mathbf v_m  \in \mathbb C^{N_P \times 1}|\|\mathbf v_m\|_2 = 1\}_{m=1}^M$ was introduced by 3GPP to enable limited feedback schemes in FDD massive MIMO wireless systems. Specifically, based on the codebook $\mathcal V$, two scalar values, the precoding matrix indicator (PMI) 
\begin{align}
    m_0(t) &= \mathop{\arg\max}_{m=1,\cdots,M} ~~\mathbf v^H_m \mathbf R(t)  \mathbf v_m\nonumber \\
    &= \mathop{\arg\max}_{m=1,\cdots,M} ~~\mathbf v^H_m  \mathbf Q(t)^H \mathbf{C} \mathbf Q(t)  \mathbf v_m,  t = 1, \cdots, T, \label{PMI}
\end{align}
and the channel quality indicator (CQI)
\begin{align}
    \eta (t) =\mathbf v^H_{m_0(t)} \mathbf Q(t)^H \mathbf{C} \mathbf Q(t)  \mathbf v_{m_0(t)}\label{CQI},~~ t = 1, \cdots, T, 
\end{align} 
are computed by the UE and then fed back to the BS.
 
In the context of system model introduced above, the CCM reconstruction challenge mentioned in Section I-A can be concretely stated as: {\it how can the BS reconstruct the downlink CCM $\mathbf C \in \mathbb C^{N_A \times N_A}$ from PMI values $\{m_0(t)\}_{t=1}^T$ and CQI values $\{\eta(t)\}_{t=1}^T$, by exploiting the structure of Type I codebook  $\mathcal V$ and the pilot weighting matrices $\{\mathbf Q(t)\}_{t=1}^T$?} 

This problem is highly under-determined and thus non-trivial. To effectively solve such a problem, the underlying principle is to embed all the {\it prior} information we have about the CCM $\mathbf C$ (to be estimated) into the problem formulation, see, e.g., \eqref{PMI} and \eqref{CQI}. Specifically, we model the feasible set of CCM estimate $\mathbf{\hat C}$ as follows:
\begin{align}
	 &\mathcal B ( \mathbf{\hat C}; \{\mathbf Q(t)\}_{t=1}^T) \label{orginal_prob_cons}  \\
	&= \Big \{\mathbf{\hat C} |~\mathbf v^H_m \mathbf Q(t)^H \mathbf{\hat C} \mathbf Q(t) \mathbf v_m \leq  \mathbf v^H_{m_0(t)} \mathbf Q(t)^H \mathbf{\hat C} \mathbf Q(t) \mathbf v_{m_0(t)}, \nonumber\\
	& ~~~~~~~~~~~~t = 1,\ldots, T,~~ \forall \mathbf v_m \in \mathcal V, \tag{C1} \\
	& ~~~~~~~~~~\mathbf v^H_{m_0(t)} \mathbf Q(t)^H \mathbf{\hat C} \mathbf Q(t) \mathbf v_{m_0(t)} = \eta(t), \nonumber \\
	& ~~~~~~~~~~~~ t = 1,\ldots, T,~~ \tag{C2} \\ 
	& ~~~~~~~~~~\mathbf{\hat C} \succeq \mathbf 0_{N_A}, \tag{C3}\\
    & ~~~~~~~~~~\mathrm{Tr}(\mathbf{\hat C} ) \leq b \tag{C4} ~~\Big \}, 
\end{align}
which comprises four groups of inequality/equality constraints (C1)-(C4). The geometries of these constraints are introduced as follows. First, the inequalities in (C1) come from the definition of PMI (see \eqref{PMI}), and indicate that the CCM estimate $\mathbf{\hat C}$ should lie in the following  polyhedron: 
\begin{align}
    \bigcap_{t=1}^T &\{\mathbf{\hat C} |~ \mathbf v^H_m \mathbf Q(t)^H \mathbf{\hat C} \mathbf Q(t) \mathbf v_m   \nonumber\\ 
    &\underbrace{~~~~~~~~~\leq ~\mathbf v^H_{m_0(t)} \mathbf Q(t)^H \mathbf{\hat C} \mathbf Q(t) \mathbf v_{m_0(t)},~~\forall \mathbf v_m \in \mathcal V}_{\triangleq \mathcal P_t}\},\label{notation_P}
\end{align}
which can be shown to be a convex cone (see Appendix \ref{Proof_poly_1}).  On the other hand, the definition of CQI (see \eqref{CQI}) suggests that $\mathbf{\hat C}$ is also in the intersection of hyper-planes defined by:
\begin{align}
   \bigcap_{t=1}^T \{ \underbrace{\mathbf{\hat C} |~ \mathbf v_{m_0(t)}^H \mathbf Q(t)^H \mathbf{\hat C} \mathbf Q(t) \mathbf v_{m_0(t)} = \eta(t)}_{\triangleq \mathcal H_t}\}. \label{notation_H}
\end{align}
The constraint (C3) encodes the positive semi-definiteness of $\mathbf{\hat C}$, which is a convex cone and is denoted by $\mathcal P_0$. Finally, without loss of generality, we introduce a trace upper bound $b$ in (C4) to eliminate the scaling ambiguity of estimation\footnote{The determination of $b$ will be introduced in the next section.}, and thus the constraint (C4) is a convex set and denoted by $\mathcal H_0$.

It can be shown (see Appendix \ref{Proof_shrinkage}) that 
\begin{align}
    \mathcal B ( \mathbf{\hat C}; \{\mathbf Q(t)\}_{t=1}^{\kappa_2}) \subseteq  \mathcal B ( \mathbf{\hat C}; \{\mathbf Q(t)\}_{t=1}^{\kappa_1}), ~~  \kappa_1 \leq \kappa_2 ,
    \label{shrinkage}
\end{align}
which suggests that incorporating additional pairs of feedback values (i.e., PMIs and CQIs) would tend to shrink the feasible set. If the feasible set is small enough (and even contains one matrix), then any element in such a set will be a good CCM estimate. This observation motivates the following problem formulation:
\begin{align}
\min_{ \{\mathbf Q(t)\}_{t=1}^T} \mathrm{Vol}( \mathcal B ( \mathbf{\hat C}; \{\mathbf Q(t)\}_{t=1}^T)). \label{orginal_prob}
\end{align}
In \eqref{orginal_prob}, $\mathrm{Vol}(\cdot)$ measures the volume of  $\mathcal B ( \mathbf{\hat C};\!  \{\mathbf Q(t)\}_{t=1}^T)$. In practice, the pilot weighting matrices $\{\mathbf Q(t)\}_{t=1}^T$ are allowed to be devised\footnote{\textcolor{black}{Though literature, e.g.,\cite{precoder1,precoder2}, provides pilot matrix design schemes, they consider the communication system with a hybrid analog-digital architecture and demand channel estimation, thus not suitable for the current case.}}. Note that in the $t$-th communication round, the feedback values (PMI $m_0(t)$ and CQI $\eta(t)$) are determined by the pilot weighting matrix $\mathbf Q(t)$, as seen in \eqref{PMI} and \eqref{CQI}. Therefore, the underlying rationale of solving problem \eqref{orginal_prob} is that matrices $\{\mathbf Q(t)\}_{t=1}^T$ should be designed such that the feedback values are the most informative ones that make the feasible set as small as possible.  

\section{CCM Reconstruction Algorithm: Development and Analysis}
\label{sec:CCM Reconstruction}

Computing the analytical expression for the volume of  $\mathcal B ( \mathbf{\hat C};\!  \{\mathbf Q(t)\}_{t=1}^T)$ is difficult (and even not tractable), thus making the associated optimization problem (10) not straightforward and challenging. In this paper, we generalize the idea of cutting plane method \cite{cp1,cp2,cp3} from seeking {\it a feasible vector} to estimating {\it a positive semi-definite CCM}, in the context of limited feedback FDD wireless systems. The generalization is non-trivial, especially when optimizing the weighting matrices $\{\mathbf Q(t)\}_{t=1}^T$ to consecutively shrink the feasible set $\mathcal B ( \mathbf{\hat C};\!  \{\mathbf Q(t)\}_{t=1}^T)$, due to a large number of inequality constraints involved. Notably, in (C1) of \eqref{orginal_prob_cons}, there are $MT $ (e.g., $256\times 10 = 2560$)\footnote{Typically, there are $M = 256$ codewords in Type I codebook. $T$ is affected by multiple factors, including the movement of UE, the change of communication environment, etc. \textcolor{black}{In general, $T$ is less than $50$ under the channel stationary scenario \cite{IOT2}.}} inequality constraints. To the best knowledge of the authors, there is still no work discussing the effective design of these weighting matrices, and we propose a viable algorithm with convergence guarantee in this section. 

\label{sec:channel_reconstruction}
\vspace{-0.2cm}
\subsection{Weighting Matrix Design} 
\label{sec:WMD}
From \eqref{orginal_prob_cons}-\eqref{shrinkage}, the feasible set $\mathcal B ( \mathbf{\hat C}; \{\mathbf{Q}(t)\}_{t=1}^T)$ can be expressed as:
\begin{align}
\mathcal B ( \mathbf{\hat C};\{\mathbf{Q}(t)\}_{t=1}^T) = \bigcap_{i=0}^T\left(\mathcal P_i \cap \mathcal H_i\right). \label{decomposition}
\end{align}
In order to minimize the volume of set $\mathcal B ( \mathbf{\hat C};\{\mathbf{Q}(t)\}_{t=1}^T)$, weighting matrices $\{\mathbf{Q}(t)\}_{t=1}^T$ are supposed to make the intersections in \eqref{decomposition} small. Since each $\mathcal H_t$ defines a hyper-plane, $\{\mathbf{Q}(t)\}_t$ should vary in different communication rounds in order to make the intersections among the hyper-planes $\{\mathcal H_t \}_t$ small, which is easy to achieve. On the other hand, since each set $\mathcal P_t$ defines a polyhedron that is composed of $M$ (e.g., $256$) hyper-planes, $\{\mathbf{Q}(t)\}_t$ should be carefully designed such that the intersections among the polyhedrons $\{\mathcal P_t \}_t$ is small, which is non-trivial. In other words, polyhedron $\mathcal P_{t+1}$ should significantly cut the intersections of previous polyhedrons $\{\mathcal P_i\}_{i=1}^t$. This coincides with the idea of ``neutral/ deep cut'' in the literature of cutting plane methods \cite{cp1}. Particularly, suppose that $\mathbf{\hat C}(t)$ is the analytical center \cite{cp2,cp3} of the feasible set $\mathcal B ( \mathbf{\hat C}; \{\mathbf{Q}(i)\}_{i=1}^t)$, it should be excluded from (or on the supporting faces of) polyhedron $\mathcal P_{t+1}$, which is determined by $\mathbf Q(t+1)$ (as defined in \eqref{notation_P}). 
To achieve this goal, we formulate the following problem: 
\begin{align}
    &\mathrm{Find}~ \mathbf Q(t+1) \nonumber \\
    \mathrm{s.t.} ~~ & \mathbf{v}_{m^{\prime} (t+1) }^{H} \mathbf{Q}(t+1)^{H} \mathbf{\hat{C}}(t) \mathbf{Q}(t+1) \mathbf{v}_{m^{\prime}(t+1)} \nonumber\\
    & ~~~\geq \mathbf{v}^H_{m_0(t+1)} \mathbf{Q}(t+1)^{H} \mathbf{\hat{C}}(t) \mathbf{Q}(t+1) \mathbf{v}_{m_0(t+1)}, \nonumber\\
    &\exists ~ m^{\prime}(t+1) \in \{1,2,\cdots, M\}. \label{WMD}
\end{align}
The challenge of solving problem \eqref{WMD} lies in that the value of  $m_0(t\!+\!1)$ has not been fed back when optimizing $\mathbf Q(t\!+\!1)$. Therefore, we need to design $\mathbf Q(t\!+\!1)$ such that the inequality in problem \eqref{WMD} holds for all the possible values of $m_0(t\!+\!1)$, making the solution of problem \eqref{WMD} irrelevant to $m_0(t\!+\!1)$. The basic idea of designing such $\mathbf Q(t\!+\!1)$ is to make $\mathbf{v}_{m^{\prime}(t+1)}$ as the first principal eigenvector of matrix $\mathbf{Q}(t\!+\!1)^{H} \mathbf{\hat{C}}(t) \mathbf{Q}(t\!+1)$ (corresponding to the largest eigenvalue), that is
\begin{align}
    \mathbf{v}_{m^{\prime} (t+1)} = \mathop{\arg\max}_{\|\mathbf{x}\| = 1} ~~\mathbf{x}^H\mathbf{Q}(t+1)^{H} \mathbf{\hat{C}}(t) \mathbf{Q}(t+1)\mathbf{x}.
\end{align}
To achieve this,  we first devise an auxiliary matrix $\mathbf{R}_{t+1}$ for different ranks of $\mathbf{\hat C}(t)$, and show that the solution of problem \eqref{WMD} can be given by the following equation: 
\begin{align}
    \mathbf{Q}(t+1)^{H} \mathbf{\hat{C}}(t) \mathbf{Q}(t+1) = \mathbf{R}_{t+1}. \label{WMD_2}
\end{align}
The key results are summarized in \textbf{Proposition \ref{Prop:1}} and \textbf{Proposition \ref{Prop:2}} as follows.

\begin{proposition}
\label{Prop:1}
If $ \mathrm{rank} (\hat{\mathbf{C}}(t)) = N_A$, construct a matrix 
\begin{align}
\mathbf{R}_{t+1} = &\sigma_{1} \mathbf{v}_{m^{\prime}(t+1)} \mathbf{v}_{m^{\prime}(t+1)}^{H} + \sum_{i=2}^{N_P}\sigma_i\mathbf{u}_{i-1}\mathbf{u}_{i-1}^H, 
\end{align}
where $m^{\prime}(t+1) \in \{1,2,\cdots,M\}$ ; hyper-parameters $\{\sigma_{n} \}_{n=1}^{N_P}$ and $\{\mathbf{u}_{n}\}_{n=1}^{N_P-1}$ are pre-selected such that  $\sigma_{1}  \geq \sigma_{2}  \geq  \cdots \geq \sigma_{N_{P}}  $ and columns $\{ \mathbf{v}_{m^{\prime}(t+1)},\mathbf{u}_{1}  ,\cdots, \mathbf{u}_{N_{P}-1}  \}$ are all orthonormal.
The solution of problem \eqref{WMD} can be provided by solving equation \eqref{WMD_2}, with the closed-form expression:
    \begin{align}
     \mathbf Q(t+1) = \hat{\mathbf{C}}(t)^{-\frac{1}{2}} \mathbf X_{t+1} \mathbf \Sigma_{t+1}  \mathbf Y_{t+1}, \label{p1_1}
    \end{align}
    where
    \begin{align}
    &\mathbf Y_{t+1}   = \left [ \mathbf{v}_{m^{\prime}(t+1)}, \mathbf{u}_{1}, \cdots,   \mathbf{u}_{N_P-1} \right],  \label{p1_2}\\
    &\mathbf \Sigma_{t+1}  = \mathrm{diag} \Big(\{ \underbrace{\sqrt{\sigma_{1}}, \sqrt{\sigma_{2}}, \cdots,  \sqrt{\sigma_{N_P}}}_{N_P}, \underbrace{0, \cdots, 0}_{N_A-N_P} \} \Big), \label{p1_3}
    \end{align}
    and  $\mathbf X_{t+1}$ is a random unitary matrix, i.e., $\mathbf X_{t+1}^H \mathbf X_{t+1}  = \mathbf I_{N_A}$.
    \begin{proof}
     See Appendix \ref{Proof_P1}.
    \end{proof}
\end{proposition}

\begin{proposition}
\label{Prop:2}
    If $ \mathrm{rank} (\hat{\mathbf{C}}(t)) = K< N_A$, construct a matrix:
      \begin{align}
    \mathbf R_{t+1} = &\sigma_{1} \mathbf{v}_{m^{\prime}(t+1)} \mathbf{v}_{m^{\prime}(t+1) }^{H}+ \sum_{i=2}^{N_P}\sigma_i\mathbf{u}_{i-1}\mathbf{u}_{i-1}^H, 
    \end{align}
where $m^{\prime}(t\!+\!1) \in \{1,2,\cdots,M\}$ and $\{\mathbf{u}_{n}\}_{n=1}^{N_P-1}$ are preselected such that $\{ \mathbf{v}_{m^{\prime}(t+1)},\mathbf{u}_{1} ,\cdots, \mathbf{u}_{N_P-1} \}$ are all orthonormal. When $K \!<\! N_P$, the hyper-parameters $\{\sigma_{n}\}_{n=1}^{N_P}$ follows 
    \begin{align}
        \sigma_{1}\! \geq \!\sigma_{2} \!\geq  \!\cdots \!\geq \!\sigma_{K}  \textgreater  \sigma_{K+1} \! = \sigma_{K+2} = \!\cdots = \!\sigma_{N_P} \!= \!0.
    \end{align}
    Otherwise, when $N_P\leq K<N_A$,  $\{\sigma_{n}\}_{n=1}^{N_P}$ need to satisfy 
    \begin{align}
        \sigma_{1} \geq \sigma_{2} \geq  \cdots \geq  \sigma_{N_P} > 0.
    \end{align}
    Then, the solution of problem \eqref{WMD} is
	
	\begin{align}
	\mathbf Q(t+1) = 
	\left[
	\begin{array}{c}
	\mathbf{U}^{-H}_{t,11} \mathrm{diag}(\mathbf{s}_t)^{-1} \mathbf{U}^{H}_{t,1}\mathbf{F}_{t+1}\\ 
	\mathbf{0} 
	\end{array}
	\right ] + \mathbf{O}_{t},  \label{p2_1}
	\end{align}
	where 
	\begin{align}
	&\mathbf{F}_{t+1}\!= \mathbf{R}^{\frac{1}{2}}_{t+1} = \mathbf X_{t+1} \mathbf \Sigma_{t+1} \mathbf Y_{t+1};\label{notation_F}\\
	&\mathbf Y_{t+1} \!= \!\left [ \mathbf{v}_{m^{\prime}(t+1)}, \mathbf{u}_{1}, \cdots,   \mathbf{u}_{N_{P}-1} \right] \in \mathbb{C}^{N_P \times N_P};\\
	&\mathbf \Sigma_{t+1} \!=
\!\!	\left[
	\begin{array}{c}
		\!\mathrm{diag}\left(\{ \!\sqrt{\sigma_{1}}, \sqrt{\sigma_{2}}, \cdots,  \sqrt{\sigma_{N_P}}\}\right) \! \in \!\mathbb{C}^{N_P\!\times\! N_P}\!\!\!\\ \hline 
		\mathbf{0} \in\mathbb{C}^{(N_A-N_P) \times N_P }
	\end{array}
	\right];\\
	&\mathbf{X}_{t+1}  \in \mathbb{C}^{N_A \times N_A} \text{ and }\nonumber\\
	&~~\mathbf X^H_{t+1} \mathbf X_{t+1} \!=\! \left[
	\begin{array}{c|c}
		\mathbf{I}_{K \times K} \in\mathbb{C}^{K\!\times\! K}& \mathbf 0 \in\mathbb{C}^{(N_A\!-\!K) \!\times\! K}\\ \hline 
		\mathbf 0 \in\mathbb{C}^{K \!\times \!(N_A\!-\!K)}& \mathbf 0 \in\mathbb{C}^{(N_A\!-\!K)\! \times\! (N_A\!-\!K)}
	\end{array}
	\right];\\
	& \mathbf{U}_{t} = 
	\begin{bmatrix}
	\begin{array}{c|c}
	\mathbf{U}_{t,1} \in \mathbb{C}^{N_A\!\times\! K}  
&	\mathbf{U}_{t,2} \in\mathbb{C}^{N_A \!\times\! (N_A\!-\!K)}
	\end{array}
    \end{bmatrix}\nonumber\\\nonumber
& ~~~~ =  \begin{bmatrix}
\begin{array}{c|c} 
  \mathbf{U}_{t,11} \in \mathbb{C}^{K\!\times\! K} &  \mathbf{U}_{t,21} \in \mathbb{C}^{K\!\times\! (N_A\!-\!K)}\\ \hline 
   \mathbf{U}_{t,12} \in  \mathbb{C}^{(N_A\!-\!K)\!\times\! K}   &    \mathbf{U}_{t,22} \in \mathbb{C}^{(N_A\!-\!K)\!\times\! (N_A\!-\!K)}
\end{array}
\end{bmatrix}\\
    &~~\text{ collects the eigenvector of }\mathbf{\hat{C}}(t)^{\frac{1}{2}};\label{notation_U}\\
	& \mathbf{s}_t = \left[ \sigma^{c}_{t,1},  \cdots, \sigma^{c}_{t,K} \right] \nonumber\\
	&~~\text{ contains all nonzero singular values of }\mathbf{\hat{C}}(t)^{\frac{1}{2}};\label{notation_diag_s}\\
	&\mathbf{O}_{t} \in  \mathrm{Null} (\mathbf{U}_{t,1}^H). \label{O_def}
	\end{align}
	
    \begin{proof}
        See Appendix \ref{Proof_P2}.     
	\end{proof}
\end{proposition}

In \textbf{Proposition \ref{Prop:1}} and \textbf{Proposition \ref{Prop:2}}, when constructing  $\mathbf R_{t+1}$, there are a number of hyper-parameters that need to be pre-defined, including $m^{\prime}(t+1)\in\{1,2,\cdots,M\}$, $\{\sigma_{n}\}_{n=1}^{N_P}$, $\{\mathbf{u}_{n}\}_{n=1}^{N_P-1}$ and unitary matrix $\mathbf{X}_{t+1}$.
Different strategies of setting those hyper-parameters result in different $\mathbf{Q}(t+1)$, which further result in different number of communication rounds for the algorithm to converge\footnote{The convergence analysis will be provided in Section \ref{sec: convergence}.}.
In Table \ref{table:WMD}, we list some strategies, and use the ``check'' symbol to mark those with the best performance in terms of both reconstruction accuracy and convergence speed in extensive simulation studies.

\begin{table*}[!tbp]
    \centering
    \caption{Different Strategies of Setting Hyper-parameters For Problem \eqref{WMD}.}
    \renewcommand{\arraystretch}{1.7}
    \begin{tabular}{|l|l|l|l|}
    \hline
    \multicolumn{1}{|c|}{\multirow{3}{*}{\centering $m^{\prime}(t+1) \in \{1,2,\cdots,M\}$}} & Random Strategy & Randomly selected from $\{1,2,\cdots, M\}$.                                                                                                                                                                                                                               &  \\ \cline{2-4}
    \multicolumn{1}{|c|}{}                                                        
    & Reuse Strategy & $ m^\prime(t+1)  = \arg\max_m \mathbf{v}_{m}^H \mathbf{Q}^H (t) \hat{\mathbf{C}}(t) \mathbf{Q}(t) \mathbf{v}_m, \forall \mathbf{v}_{m} \in \mathcal{V}.$                                                                                         &  \\ \cline{2-4} 
    \multicolumn{1}{|c|}{}                                                        &Mixture Strategy &
    \begin{tabular}[c]{@{}l@{}}  Define $\gamma_t = \sqrt{\frac{\|\hat {\mathbf C}(t-1)-\hat{\mathbf{C}}(t)\|^2}{N_A^2}}$. If $\gamma_t \geq \epsilon$, random strategy; otherwise reuse strategy.\end{tabular}                                                                      &   \checkmark  \\ \hline
    \multirow{2}{*}{$\{\sigma_{n}\}_{n=1}^{N_P}$}                            &Equality Strategy & $\sigma_{1} = \sigma_{2} = \cdots = \sigma_{N_P} =1$.                                                                                                                                                                                                      & \checkmark  \\ \cline{2-4}              & Sampling-sorting Strategy & Sample each  $\sigma_{n} \sim \mathrm{U}(0,1)$  and then sort these values             &  \\ \hline
    \multirow{2}{*}{$\mathbf{X}_{t+1}$}                                             &Designed Strategy & \begin{tabular}[c]{@{}l@{}}Let $\mathbf{B}(t)=\sum_{i=1}^{t} \mathbf{Q}(i) \exp\{10(i-t)\}$ .\\ $ \mathbf{X}_{t+1} = \arg \min \operatorname{Tr}\left(\mathbf{Q}^{H}(t+1) \mathbf{B}(t)+\mathbf{B}^{H}(t) \mathbf{Q}(t+1)\right)$ (see Appendix \ref{History_Q_info}).\end{tabular} &  \checkmark\\ \cline{2-4} 
                 & Random Strategy & Randomly generated.            &   \\ \hline
    \end{tabular}
    \label{table:WMD}
    
\end{table*}

\subsection{Analytic Center Acquisition}
\label{sec:CMO}
In problem \eqref{WMD}, the analytical center matrix  $\mathbf{\hat{C}}(t)$ of the feasible set $\mathcal B ( \hat{\mathbf C}; \{\mathbf{Q}(i)\}_{i=1}^{t})$ is required. Inspired by the analytical center optimization in the framework of cutting plane method \cite{cp2,cp3}, given the weighting matrices $\{\mathbf{Q}(i)\}_{i=1}^{t}$, we propose to find the center matrix $\mathbf{\hat{C}}(t)$ via solving the following problem:
\begin{align}
&\max_{ {\mathbf C} } \sum_{i=1}^{t}\sum_{m=1}^M \frac{1}{\eta(i)} \log \Big ( \mathbf v_{m_0(i)}^H \mathbf Q(i)^H {\mathbf C} \mathbf Q(i) \mathbf v_{m_0(i)} \nonumber\\
&~~~~~~~- \mathbf v_m^H \mathbf Q(i)^H {\mathbf C} \mathbf Q(i) \mathbf v_m \Big )+ \log \det ( {\mathbf C} ) - \lambda ||{\mathbf C}||_{*}\nonumber \\
&~~\mathrm{s.t.} ~~ \mathbf v^H_{m_0(i)} \mathbf Q(i)^H {\mathbf C} \mathbf Q(i) \mathbf v_{m_0(i)} = \eta (i), ~~i = 1,\ldots, t,\nonumber\\
&~~~~~~~~ a \leq \mathrm{Tr}({\mathbf C} ) \leq b, \label{CMO}  
\end{align}
where $|| \cdot ||_{*}$ denotes the nuclear norm of the argument, and $\lambda\geq 0$ is the regularization parameter. 
In addition, $a$ and $b$ (see (C4) in \eqref{orginal_prob_cons}) are the lower bound value and the upper bound value of the trace of the ground-truth CCM $\mathbf C$, respectively. 

Note that problem \eqref{CMO} is different from the standard formulation of analytic center acquisition problem \cite{cp2,cp3}, due to the incorporation of the system model information (see Section \ref{sec:system}) for more accurate CCM estimation. Specifically, the low-rank structure of $\hat {\mathbf C}(t)$ is promoted by the added nuclear norm based regularization term. The inequality constraint $\mathrm{Tr}(\mathbf{C})\geq a$ is introduced to further reduce the feasible region. CQI values $\{\eta(i)\}_{i=1}^t$ are employed to re-weight each log-barrier term to make $\hat {\mathbf C}(t)$ closer to the ground-truth $\mathbf{C}$. Even with these differences, problem \eqref{CMO} is still convex and can be well solved by the CVX solver\footnote{http://cvxr.com/cvx/.}. 

For the hyper-parameters of problem \eqref{CMO},  the regularization parameter $\lambda$ needs to be tuned,  while the upper/lower bound values $\{a ,b\}$ can be analytically obtained. 

\subsubsection{Strategy for setting $b$}
\label{subsubsection:strategy_b}
As the upper bound of $\mathrm{Tr}(\mathbf C)$, the value of $b$ depends on the normalization scheme adopted for the ground-truth $\mathbf C$ at the UE side. If trace normalization is utilized, then $b=1$ is an appropriate choice. If Frobenius normalization is used, then $b$ should take value $N_U$. \textcolor{black}{If the normalization scheme is unknown at the BS, we provide one heuristic scheme to estimate $b$ by exploiting the available historical CCM (See Appendix \ref{Strategy_b}).}


\subsubsection{ Strategy for setting $a$}
For the lower bound $a$, it can be given by the following proposition.
\begin{proposition}
	Given CQI $\{\eta (t)\}^{T}_{t=1}$, the trace of the ground-truth CCM $\mathbf C$ satisfies
	\begin{align}
	\mathrm{Tr}(\mathbf C) \geq \max_{t=1,\cdots,T} \frac{\eta(t)}{\mathrm{Tr}\left(\mathbf{Q}(t) \mathbf{v}^{}_{m_0(t)} \mathbf{v}_{m_0(t)}^{H} \mathbf{Q}^{H}(t)\right)}.
	\end{align}
	\begin{proof}
		See Appendix \ref{Strategy_a}.
	\end{proof}
	\label{Prop:3}
\end{proposition}
 
\subsection{Algorithm Summary}
 
\begin{algorithm}[!tbp]
\caption{\bf: The Proposed Channel Covariance Matrix  Reconstruction Algorithm}
\noindent {\bf Initialization:} Set initial value of $\mathbf {Q}(1)$. 

For the $t$-th communication round  ($t \geq 1$):

$~~$\noindent Update center matrix $\hat{\mathbf C}(t)$ via solving \eqref{CMO};

$~~$\noindent If $\mathrm{rank}(\hat{\mathbf C}(t)) = N_A$, update $\mathbf {Q}(t+1)$ using \eqref{p1_1};

$~~$\noindent If $\mathrm{rank}(\hat{\mathbf C}(t)) < N_A$, update $\mathbf {Q}(t+1)$ using \eqref{p2_1};

\noindent {Until the communication round $T$.}

\noindent {\bf Output:} Channel covariance matrix estimate $\hat{\mathbf C}_T$.
\label{alg:cutting_plane}

\end{algorithm}
 
From previous subsections, it can be seen that the design of weighting matrix $\mathbf Q(t+1)$ requires the analytical center $\mathbf{\hat C}(t)$, while the computation of $\mathbf{\hat C}(t)$ needs a set of weighting matrices $\{\mathbf Q(i)\}_{i=1}^{t}$. This motivates the alternating updates of center CCM matrix $\mathbf{\hat C}(t)$ and weighting matrix $\mathbf Q(t+1)$, as summarized in {\bf Algorithm \ref{alg:cutting_plane}}. The algorithm is composed of two steps: weighting matrix design (see Section \ref{sec:WMD}) and analytic center acquisition (see Section \ref{sec:CMO}). In the $t$-th communication round ($t \geq 1$), the analytic center acquisition provides a feasible CCM estimate $\hat{\mathbf C}(t) \in \mathcal B ( \mathbf{\hat C}; \{\mathbf{Q}(i)\}_{i=1}^t)$, based on which weighting matrix $\mathbf{Q}(t+1)$ is designed to achieve a ``neutral/deep'' cut for the set $\mathcal B ( 
\mathbf{\hat C}; \{\mathbf{Q}(i)\}_{i=1}^{t+1})$. As shown in the next subsection, when $T$ is large enough, the set $\mathcal B ( \mathbf{\hat C}; \{\mathbf{Q}(i)\}_{i=1}^{T})$ will be very small such that any element in such a set provides a good CCM estimate.

\subsection{Convergence Analysis}
\label{sec: convergence}
In this section, we theoretically show that the CCM estimate sequence $\{\hat{\mathbf C}(t)\}_{t=1}$, in which $\hat{\mathbf C}(t)$ is the solution of problem \eqref{CMO} in the $t$-th communication round, will converge to the ground-truth CCM $\mathbf{C}^*$ when $t$ goes to infinity. To show this, we first present the following property that characterizes the feasible set of problem \eqref{CMO}.

\begin{property}
\label{property:1}
In the $t$-th communication round, denote the feasible set of problem \eqref{CMO} as $\mathcal{F}_t, ~t = 1,2,\cdots,T$, we have:
\begin{enumerate} 
    \Item [{\it  i)}]
    \begin{align}
    \mathcal{F}_T \subseteq \mathcal{F}_{T-1} \subseteq \cdots \subseteq \mathcal{F}_{1}. 
\end{align}
\Item[{\it ii)}]\begin{align}
    \mathbf{C}^* \in \mathcal{F}_t, ~~\forall t.
\end{align}
\end{enumerate}

\vspace{-0.3cm}
\begin{proof}
       see Appendix \ref{property1}.
\end{proof}
\end{property}

\textbf{Property \ref{property:1}} points out that the feasible set $\mathcal{F}_t$ will tend to shrink as the communication round $t$ increases, and always contain the ground-truth CCM $\mathbf{C}^*$. Therefore, when $t$ goes to infinity, if the set $\mathcal{F}_t$ only contains one feasible CCM $\mathbf{C}_t$, this CCM must be the ground-truth $\mathbf{C}^*$. In the following, we show the statement above indeed holds.

To facilitate the analysis, we introduce a relaxed version of problem \eqref{CMO} as follows:
\begin{align}
&\max_{ {\mathbf C} } \sum_{i=1}^{t}\sum_{m=1}^M \frac{1}{\eta(i)} \log \Big ( \mathbf v_{m_0(i)}^H \mathbf Q(i)^H {\mathbf C} \mathbf Q(i) \mathbf v_{m_0(i)} \nonumber\\
&~~~~~~~~~- \mathbf v_m^H \mathbf Q(i)^H {\mathbf C} \mathbf Q(i) \mathbf v_m \Big ) + \log \det ( {\mathbf C} ) - \lambda ||{\mathbf C}||_{*}\nonumber \\
&~~\mathrm{s.t.} ~~ \mathrm{Tr}({\mathbf C} ) \leq b.\label{relaxtion_problem}
\end{align}
By noticing that several constraints in problem \eqref{CMO} have been eliminated, it is easy to see that the feasible set $\hat{\mathcal{F}}_t$ of problem \eqref{relaxtion_problem} contains the feasible set $\mathcal{F}_t$ of problem \eqref{CMO}, i.e., $\mathcal{F}_t \subseteq \hat{\mathcal{F}}_t$.  Therefore, if set $\hat{\mathcal{F}}_t$ can be shown to monotonically shrink as $t$ goes larger, and always contains the ground-truth CCM $\mathbf{C}^*$, so does the subset $\mathcal{F}_t$. To show this, we first present the property that characterizes the set $\hat{\mathcal{F}}_t$.

\begin{property}
\label{property:2}
In the $t$-th ($t>0$) communication round, for the feasible set $\hat{\mathcal{F}}_t$ of problem \eqref{relaxtion_problem}, we have
\begin{enumerate}
    \Item [{\it i)}] 
    \begin{align}
     \mathcal{\hat F}_t = \mathcal{\bar F}_0 \cap \mathcal{\bar F}_t,
\end{align}
where 
\begin{align}
    \mathcal{\bar F}_0 = &\{\mathbf{C}| \mathrm{Tr}({\mathbf C} ) \leq b\}\cap \{\mathbf{C}|\mathbf{C} \in \mathbb{S}_+\},\\
    \mathcal{\bar F}_t = &\!\bigcap_{i=1}^t\! \bigcap_{m=1}^M \!\!\Big \{\mathbf{C}| \mathbf v_m^H \mathbf Q(i)^H {\mathbf C} \mathbf Q(i) \mathbf v_m \nonumber \\
    &~~~~~~~~~~~~~\leq \mathbf v_{m_0(i)}^H \mathbf Q(i)^H {\mathbf C} \mathbf Q(i) \mathbf v_{m_0(i)}\Big \};
\end{align}
\Item [{\it ii)}] \begin{align}
    \mathcal{\hat F}_t \subseteq \mathcal{\bar F}_t;
\end{align}
\Item [{\it iii)}] \begin{align}
    \mathbf{C}^* \in \bar{\mathcal{F}}_t.
\end{align}
\end{enumerate}
 


\end{property}

\textbf{Property \ref{property:2}} indicates that the volume of $\hat{\mathcal{F}}_t$ is up to that of $\bar{\mathcal{F}}_t$. Then inspired by \cite{nesterov}, we introduce a measure to assess its volume.

\begin{definition} Given the interior point $\mathbf{\tilde C}_t$ of set $\bar{\mathcal{F}}_t$ obtained via solving problem \eqref{relaxtion_problem}, we define the volume measure of set $\bar{\mathcal{F}}_t$ as follows:
\begin{align}
    \max_{\mathbf{C} \in \bar{\mathcal{F}}_t} \mu_t(\mathbf{C}), 
\end{align}
where
\begin{align}
   &\mu_t(\mathbf{C}) = \frac{1}{S_t}\sum_{i=1}^{t}\sum_{m=1}^M \omega_{i,m,t}\langle \mathbf{G}_{m,i}, \mathbf{C}\rangle_F; \label{notation_mu}\\
   &\mathbf{G}_{m,i}  \!=\! \mathbf{Q}(i)\mathbf{v}_{m_0(i)}\mathbf{v}^H_{m_0(i)}\mathbf{Q}^H(i)-\mathbf{Q}(i)\mathbf{v}_m\mathbf{v}^H_m\mathbf{Q}^H(i); \label{notation_gmt} \\
    &\langle \mathbf{G}_{m,i},\mathbf{ C}\rangle_F = \mathrm{Tr}\left(\mathbf{G}_{m,i}\mathbf{ C}\right); \label{notation_slack}\\
    &\omega_{i,m,t} = \frac{1}{\eta(i)\langle \mathbf{G}_{m,i},\mathbf{\tilde C}_t\rangle_F}; \label{notation_w} \\
     &S_t = \sum_{i=1}^{t}\sum_{m=1}^M \omega_{i,m,t}. \label{notation_S}
\end{align}

\label{def:notations_2}
\end{definition}

This definition extends the notion of volume measure from a set containing all vectors \cite{nesterov} to a set involving CCMs. The physical meaning of \eqref{notation_mu} can be interpreted as follows. First, equations \eqref{notation_gmt} and \eqref{notation_slack} compute the following term:
\begin{align}
     \mathbf v_{m_0(i)}^H \mathbf Q(i)^H {\mathbf C} \mathbf Q(i) \mathbf v_{m_0(i)} - \mathbf v_m^H \mathbf Q(i)^H {\mathbf C} \mathbf Q(i) \mathbf v_m, 
\end{align}
which measures the distance between  $\mathbf{C}$ to the $m$-th hyper-planes in the $t$-th communication round. Second, $\omega_{i,m,t}$ is the weighting coefficient, which is adopted to eliminate the scaling difference of $\{\langle \mathbf{G}_{m,i}, \mathbf{C}\rangle_F\}_{m=1,i=1}^{M,t}$ induced by the different $\mathbf{Q}(i)$ and $\mathbf{v}_{m_0(i)}$. Finally, $S_t$ is used for normalization. 
\vspace{0.2cm}

Using \textbf{Definition \ref{def:notations_2}}, we can show the following lemma:
\begin{lemma}
\label{Lemma:1}
For all $\mathbf{C} \in \{\mathbf{C}| \mathrm{Tr}({\mathbf C} ) \leq b\}$, it holds 
\begin{align}
    \mu_t(\mathbf{C}) \leq \frac{\lambda \langle\mathbf{I},\mathbf{C}\rangle_F}{S_t}, \label{mu}
\end{align}
where $S_t \rightarrow \infty$ as $t \rightarrow \infty$. 
\begin{proof}
       see Appendix \ref{Covergence}.
\end{proof}
\end{lemma}
\textbf{Lemma \ref{Lemma:1}} shows that the volume measure $\mu_t$ is upper-bounded. Since $S_t$ is a monotonically increasing function of $t$, as $t$ goes to infinity, the right-hand side of \eqref{mu} will become zero.
In addition, we have \textbf{Lemma \ref{Lemma:2}} stated as follows:

\begin{lemma}
\label{Lemma:2}

 For any feasible CCM $\mathbf{C} \neq \mathbf{C}^* \in \mathcal{\bar F}_t$, it will be excluded from the feasible set after the $\tau$-th ($\tau > t$) communication round, i.e., $\mathbf{C} \notin \mathcal{\bar F}_i, \forall i \geq \tau$. 
\begin{proof}
 see Appendix \ref{Convergence_2}.
\end{proof}
\end{lemma}
 {\bf Lemma \ref{Lemma:2}} guarantees that the feasible set $\mathcal{\bar F}_t$ will shrink to the set that only contains one feasible CCM $\mathbf{C}^*$.
Using these two lemma, we can show the following theorem.

\begin{theorem}
As $t \rightarrow \infty$, the feasible set $\mathcal{F}_t$  of problem \eqref{CMO} only contains one feasible CCM $\mathbf{C}^*$.
\begin{proof}
 see Appendix \ref{convergence_theorem}.
\end{proof}
\label{theorem:1}
\end{theorem}
 \textbf{Theorem \ref{theorem:1}} shows that the proposed algorithm will converge to the ground-truth $\mathbf{C}^*$ when $t$ goes to infinity. 

\noindent \emph{Remark 1 (Complexity Analysis):} The complexity of \textbf{Algorithm \ref{alg:cutting_plane}} is dominated by the cost of primal-dual interior method that solves the convex optimization problem \eqref{CMO}. In \cite{pdim}, it shows the iteration complexity of primal-dual interior method is $O(\sqrt{n}L)$ in the worst case, where $n$ is the number of variables and $L$ is the number of constraints. Therefore, it can be concluded that the proposed algorithm is with complexity order $O(\sqrt{n}L)$.

\section{Numerical Results and Discussions}
\label{sec:numerical}

In this section, numerical results are presented to showcase the effectiveness of the proposed CCM reconstruction algorithm. (i.e., {\bf Algorithm \ref{alg:cutting_plane}}).  Consider a BS with $N_A = 32$ antennas and $N_P = 8$ ports, serving a UE with $N_U = 2$ antennas. The ground-truth CCMs are provided by the channel samples generated from  QuaDRiGa. Particularly, we set the center frequency of the downlink channel to $1.275~\text{GHZ}$, and assume that the speed of the UE terminal is $3$ km/h. Type I codebook is generated according to 5G NR standards \cite{5_5G}, \cite{code1}. Each point in the following figures is an average of $100$ Monte-Carlo trials. 

The CCM reconstruction performance is measured by the root mean square error (RMSE) in $\mathrm{dB}$, i.e., $10\log_{10}\left(\sqrt{\frac{|| \mathbf C^* - \hat{\mathbf C} ||_F^2}{N_A^2}}\right)$, where $\hat{\mathbf C}$ is the reconstructed CCM and $\mathbf C^*$ is the ground-truth CCM. In addition, consider that beamforming vector $\mathbf{w}$ can be acquired via computing the first principal eigenvector of $\hat{\mathbf C}$, we also use the beam precision, which is defined as $\frac{\mathbf{w}^H{\mathbf C^*}\mathbf{w}}{d}$ (where $d$ is the largest eigenvalue of ${\mathbf C^*}$), to see how the CCM reconstruction helps the beamforming. 

In Appendix \ref{strategies_table}, the RMSEs and beam precision of the proposed algorithm under different hyper-parameter setting strategies  (see Table I) are presented, from which we can identify the practically useful strategies of hyper-parameter setting. Particularly, in weighting matrix design,
$m'(t+1)$ is randomly selected from $\{1, 2, \cdots, M \}$ if the RMSE $10\log_{10}\left(\sqrt{\frac{\|\hat {\mathbf C}(t-1)-\hat{\mathbf{C}}(t)\|^2}{N_A^2}}\right) \geq \epsilon$ (e.g., $-20 \mathrm{dB}$), otherwise, $m'(t+1)$ is obtained via solving 
\begin{align}
    m^\prime(t+1)  = \mathop{\arg\max}_{m=1,\cdots,M}~~ \mathbf{v}_{m}^H \mathbf{Q}^H (t) \hat{\mathbf{C}}(t) \mathbf{Q}(t) \mathbf{v}_m. 
\end{align}
Moreover, $\mathbf X_{t+1}$ is acquired via solving 
\begin{align}
 \mathbf{X}_{t+1}\! = \!\arg \min \mathrm{Tr}\left(\mathbf{Q}^{H}(t\!+\!1) \mathbf{B}(t)\!+\!\mathbf{B}^{H}(t) \mathbf{Q}(t\!+\!1)\right),
\end{align}
where $\mathbf{B}(t)=\sum_{i=1}^{t} \mathbf{Q}(i) \exp\{10(i-t)\}$; and $\sigma_1 = \sigma_2 = \cdots = \sigma_{N_p} = 1$. In analytic center acquisition problem \eqref{CMO}, \textcolor{black}{the upper bound $b$ is set as $1$ with the assumption that CCM is normalized by its trace (see Section \ref{subsubsection:strategy_b})}, and the lower bound $a$ follows the update equation in \textbf{Proposition \ref{Prop:3}}. 

\begin{figure}[!tbp]
	\centering
	\includegraphics[width= 3.5 in]{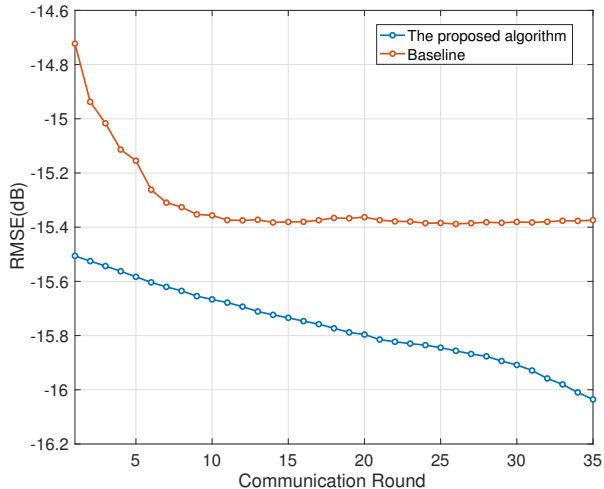}
	\caption{The RMSEs of CCM reconstruction versus communication rounds ($N_A=32, N_P = 8, N_U = 2$).}
	\label{fig:RMSE_NU2}
  \vspace{-0.3cm}
\end{figure}

The RMSEs of the proposed algorithm versus different communication rounds are presented in Fig. \ref{fig:RMSE_NU2}, with the algorithm in \cite{patent} serving as the benchmark (labeled as baseline)\footnote{\textcolor{black}{The code can be found via the link: https://github.com/wamcs/CCM-Reconstruction/blob/master/baseline.m.}}, which has been adopted in real-world 5G systems. 
In particular, the baseline algorithm estimates the CCM via the following equation:
\begin{align}
    \mathbf{\hat C} = \frac{1}{T}\sum_{t=1}^T \eta(t) \mathbf{Q}(t)\mathbf{v}_{m_0(t)} \mathbf{v}^H_{m_0(t)}\mathbf{Q}^H(t),
\end{align}
where each $ \mathbf{Q}(t)\mathbf{v}_{m_0(t)}$ can be roughly viewed as an approximation of the first principal eigenvector of the ground-truth CCM\footnote{\textcolor{black}{Specifically, $\mathbf{Q}(t)$ is designed as $\mathbf{Q}(1)\mathbf{M}_t$. Here, $\mathbf{Q}(1)$ is a matrix selected from a pre-defined set of weighting matrices $\mathcal{Q}$ (which can be found in our open-source code, see link: https://github.com/wamcs/CCM-Reconstruction/blob/master/Q0.xlsx), and $\mathbf{M}_t$ is chosen from a mutually unbiased bases (MUB) codebook $\mathcal{M}$ (specified in the patent: https://patents.google.com/patent/WO2017206527A1/en}).}.
It is clear that in each communication round, the proposed algorithm gives a much lower RMSE than the benchmarking algorithm. This shows the effectiveness of the proposed principled problem formulation, which utilizes all the system information to reconstruct CCM. In contrast, the baseline algorithm is rather heuristic. It did not take CCM reconstruction as an explicit optimization objective. As a consequence,  RMSEs stop decreasing when the communication round is larger than $10$ in Fig. \ref{fig:RMSE_NU2}.

\begin{figure}[!tbp]
	\centering
	\includegraphics[width= 3.4in]{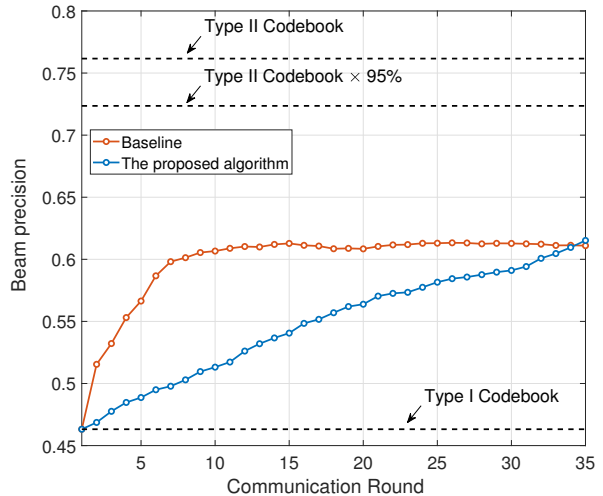}
	\caption{The beam precisions versus communication rounds ($N_A=32, N_P = 8, N_U = 2$).}
	\label{fig:Beam_Precision_NU2}
  \vspace{-0.5cm}
\end{figure}

On the other hand, we present the performance measured by beam precision in Fig. \ref{fig:Beam_Precision_NU2}. In addition to the benchmarking algorithm \cite{patent}, the beam precision using Type I codebook and Type II codebook (see the 3GPP standard \cite{code1}) also are employed as the benchmark. It can be observed that the proposed algorithm did not offer satisfactory beamforming performance unless the communication round is larger than $35$. This is because the proposed algorithm aims at reconstructing the CCM, while not its first principal eigenvector. Therefore, even though the RMSE of CCM reconstruction continues to decrease, as shown in Fig. \ref{fig:RMSE_NU2}, the first principal eigenvector of the ground-truth CCM did not get well recovered. 


\begin{figure}[!tbp]
	\centering
	\includegraphics[width= 3.5 in]{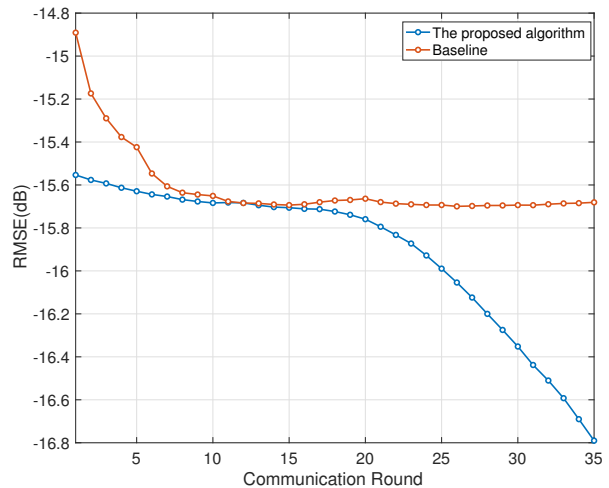}
	\caption{The RMSEs of CCM reconstruction versus communication rounds ($N_A=32, N_P = 8, N_U = 1$).}
	\label{fig:RMSE_NU1}
  \vspace{-0.5cm}
\end{figure}

\begin{figure}[!tbp]
	\centering
	\includegraphics[width= 3.5 in]{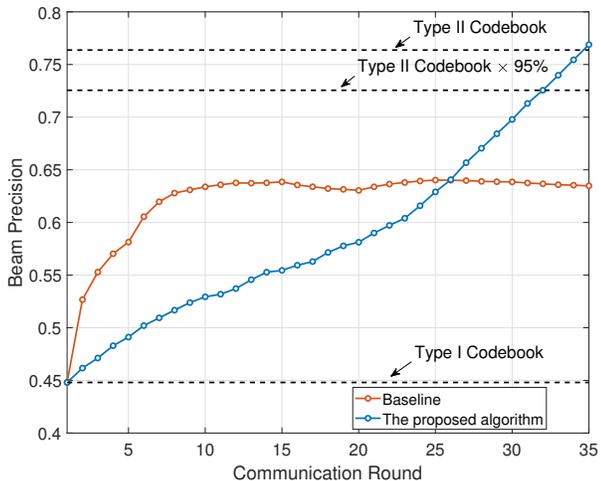}
	\caption{The beam precisions versus communication rounds ($N_A=32, N_P = 8, N_U = 1$).}
	\label{fig:Beam_Precision_NU1}
  \vspace{-0.3cm}
\end{figure}

However, if $N_U = 1$ (the UE is equipped with a single antenna), the CCM reconstruction is equivalent to the first principal eigenvector estimation of CCM. Under this setting, we present the RMSEs and beam precision of different algorithms in Fig. \ref{fig:RMSE_NU1} and Fig. \ref{fig:Beam_Precision_NU1} respectively. It is clear the proposed algorithm still offers much lower RMSEs than those of the 
baseline algorithm \cite{patent}, see Fig. \ref{fig:RMSE_NU1}. On the other hand, under the rank-1 CCM setting, the proposed CCM reconstruction algorithm equivalently seeks the optimal first principal eigenvector of $\mathbf{C}$. As a result, in Fig. \ref{fig:Beam_Precision_NU1}, the beam precision of the proposed algorithm continues to increase and exceeds those of the baseline algorithm \cite{patent} after the $26$-th communication round. Meanwhile, it touches the $95\%$ of Type II codebook based beam precision at the $32$-th communication round. This shows the excellent performance of the proposed algorithm for the UE with $N_U = 1$ antenna.

\textcolor{black}{In addition, the numerical results presented in Fig. \ref{fig:RMSE_NU2} to Fig. \ref{fig:Beam_Precision_NU1} show that the beam precision experiences a significant improvement, while the RMSE of CCM only shows a slight gain of approximately 1 dB. Notice that even when the RMSE between two matrices is small, their first principal eigenvectors can still differ significantly, especially when the element values of these matrices are small. In our specific case, the ground-truth CCM generated by QuaDRiGa has small element values on the order of $1e-3$. Therefore, even if the RMSE of the CCM only shows a slight improvement, the estimated first principal eigenvectors can exhibit significant differences, leading to a notable improvement in beam precision.}

\begin{figure}[!tbp]
	\centering
	\includegraphics[width= 3.5 in]{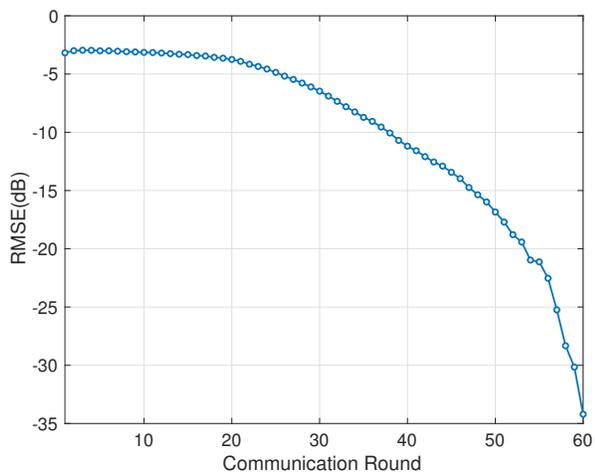}
	\caption{The RMSEs of CCM reconstruction versus communication rounds ($N_A=8, N_P = 2, N_U = 1$).}
	\label{fig:convergence}
  \vspace{-0.3cm}
\end{figure}

Finally, in Fig. \ref{fig:convergence}, the convergence of the proposed algorithm is verified under the setting: $N_U = 2, N_A = 8$ and $N_P = 2$, for the ease of illustration. As shown in Fig. \ref{fig:convergence}, the RMSE of the proposed algorithm keeps decreasing as the communication round increases. In particular, after the $60$-th communication round, the RMSE is less than $-33 \mathrm{dB}$ ($\approx 5\times 10^{-4}$), which shows the nearly exact recovery of the ground-truth CCM.

\begin{figure}[!tbp]
\vspace{-0.2cm}
	\centering
	\includegraphics[width= 3.5 in]{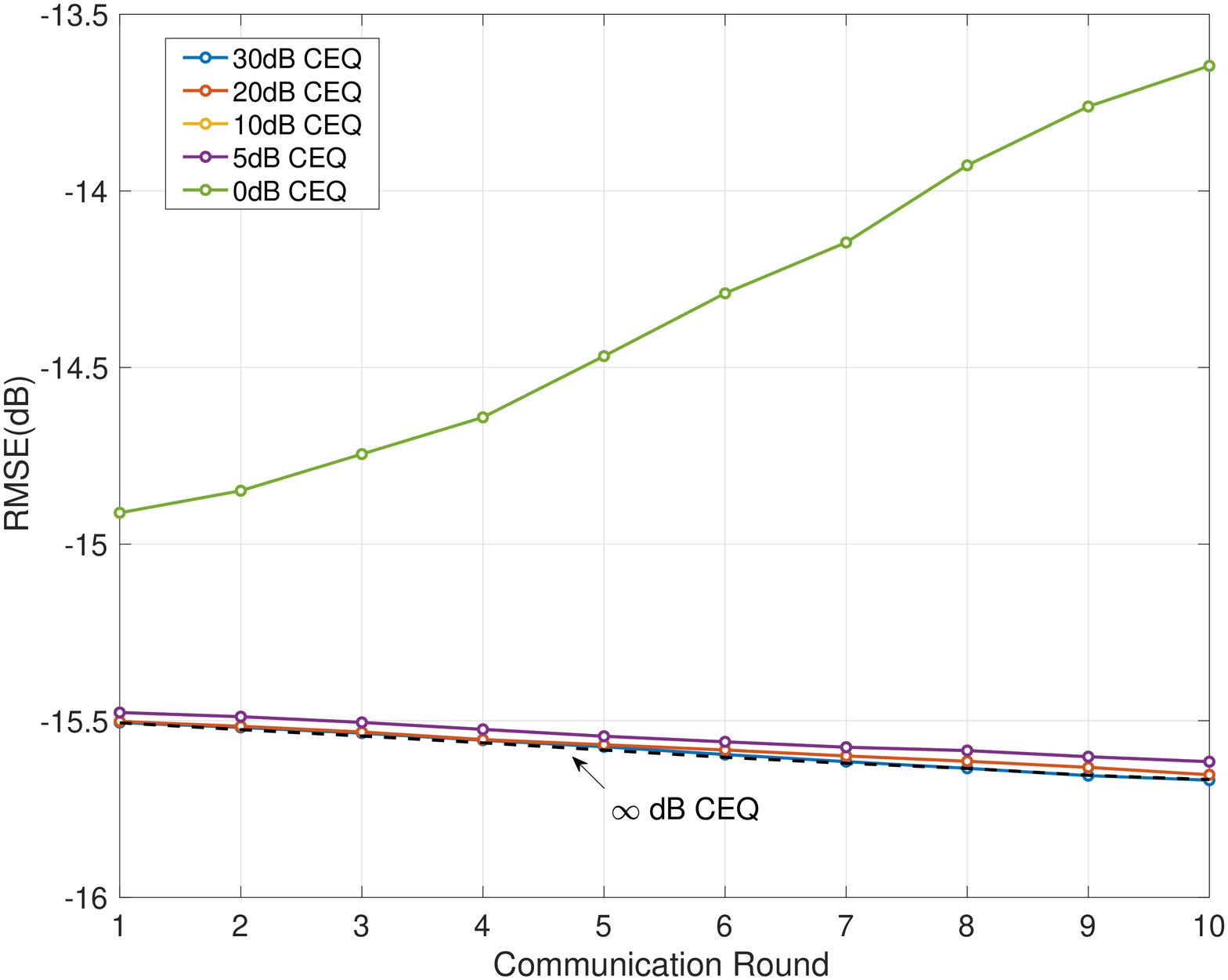}
	\caption{The RMSEs of CCM reconstruction versus communication rounds with different CEQs ($N_A=32, N_P = 8, N_U = 2$).}
	\label{fig:RMSE_NU2_ROBUST}
  \vspace{-0.3cm}
\end{figure}

\begin{figure}[!tbp]
	\centering
	\includegraphics[width= 3.5 in]{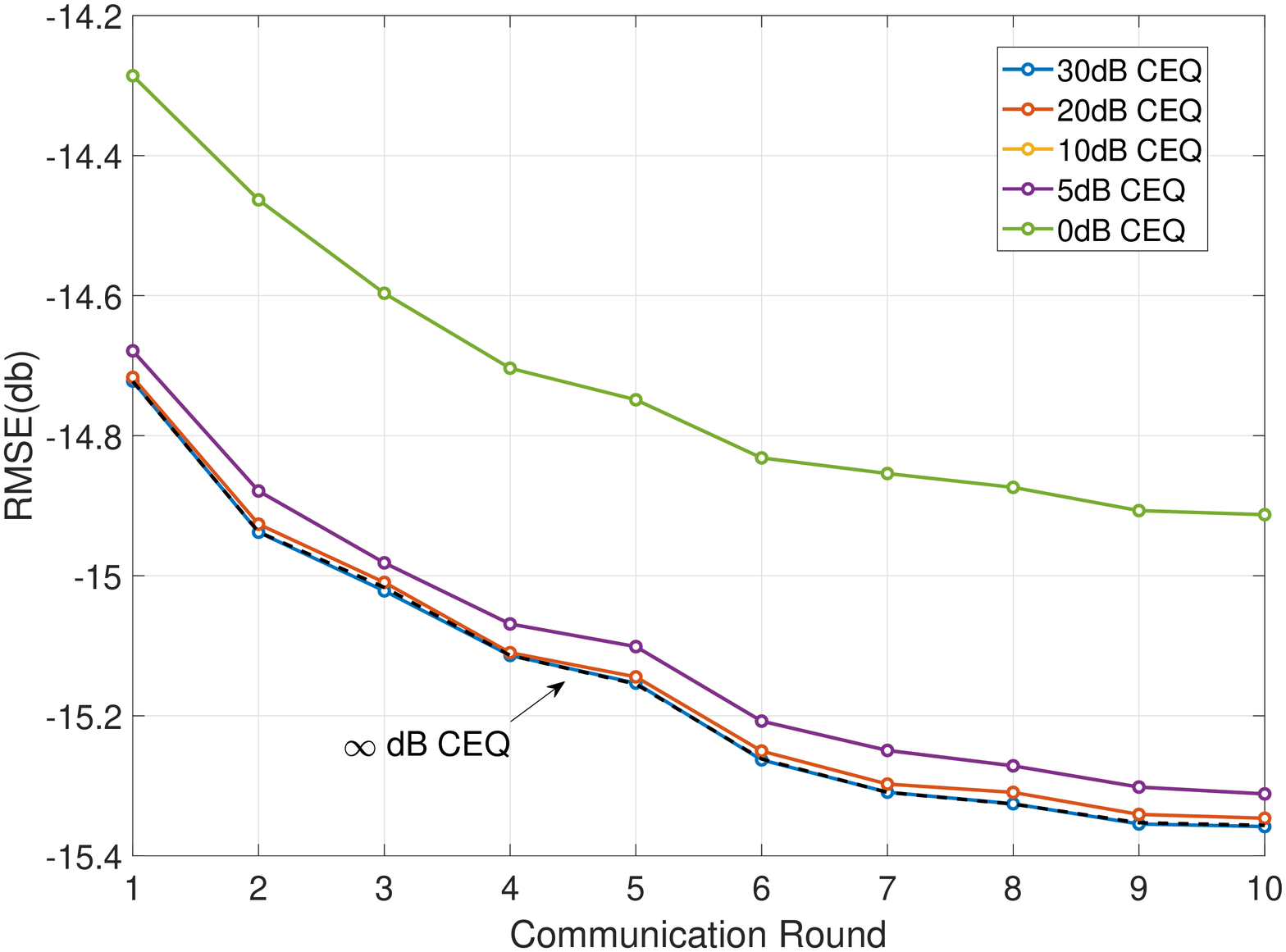}
	\caption{The RMSEs of CCM reconstruction versus communication rounds with different CEQs ($N_A=32, N_P = 8, N_U = 2$).}
	\label{fig:RMSE_NU2_PATENT_ROBUST}
  \vspace{-0.3cm}
\end{figure}

\textcolor{black}{It is worth noticing the above numerical results are under the assumption that effective channel estimate is accurate, i.e., no estimation error in $\mathbf{H}_e(t)=\mathbf{H}(t) \mathbf{Q}(t)+\mathbf{E}(t)$ (see \eqref{effective_channel}). Such an assumption is reliable since the proposed algorithm and the benchmarking algorithm are robust against receiver-side CCM estimation error. To measure the receiver-side CCM estimation performances, the channel estimation quality (CEQ) defined as
\begin{align}
  \mathrm{CEQ}(\mathbf{E}(t))=10\log_{10}\left(\frac{|\mathbf{H} (t )\mathbf{Q} (t)|_F^2}{\sigma_{e_t}}\right) \mathrm{dB}
\end{align}
(similar to the definition of SNR) is adopted. If CEQ is large, the channel is estimated accurately, and vice versa. In Fig. \ref{fig:RMSE_NU2_ROBUST}, we compare the CCM reconstruction performance (in terms of RMSEs) versus communication rounds under different CEQs. The RMSEs under $\mathrm{CEQ} = \infty \mathrm{dB}$ are plotted as the genie-aided benchmark (see the black dashed line), which corresponds to the case that there is no CCM estimation error. It can be observed that when CEQ is larger than 5dB, the RMSEs of reconstructed CCMs are similar to the genie-aided one (under $\mathrm{CEQ} =\infty \mathrm{dB}$), showing that the performance of the proposed approach is not sensitive to the CCM estimation errors in a wide range ($\mathrm{CEQ} \geq 5 \mathrm{dB}$, which is easy to achieve in practice). However, under very poor channel estimation, e.g., $\mathrm{CEQ} = 0 \mathrm{dB}$, the proposed method fails to work (see the green line). On the other hand, the benchmarking method shows similar robustness to the proposed method in Fig. \ref{fig:RMSE_NU2_PATENT_ROBUST}. }


\section{Conclusion and Future Direction}
\label{sec:conclusion}
In this paper, the reconstruction of CCM from a few feedback values at BS was investigated in 5G NR FDD massive MIMO wireless systems. Particularly, using Type I codebook, the downlink CCM reconstruction problem was formulated in a principled way by leveraging the structure of codebook and feedback values. 
The proposed effective algorithm extends the idea of cutting plane method to tackle the complicated feasible set of CCMs, and consists of two alternating steps. One is to optimize pilot weighting matrix such that the feasible set can be consecutively reduced, and another is to obtain the analytical center of  feasible set. The convergence of the proposed algorithm was theoretically analyzed. Extensive simulation results have shown the excellent performance of the proposed algorithm in terms of CCM reconstruction. In addition, when the UE is with a single antenna, a notable beamforming performance of the proposed algorithm was observed. 

\textcolor{black}{This paper only considers the single-user case, while the CCM estimation scheme in multi-user systems (e.g., massive access\cite{ke2020compressive}) is also a promising future research direction.} 


\appendix

    \subsection{The Convex Cone Property of \eqref{notation_P}}
    \label{Proof_poly_1}
    See Section \ref{subsec:1} of the supplementary document.
   
\subsection{The Shrinkage of $\mathcal B ( \mathbf{\hat C}; \{\mathbf Q(t)\}_{t=1}^T)$}
\label{Proof_shrinkage}
  See Section \ref{subsec:2} of the supplementary document.
 
\subsection{The Proof of \textbf{Proposition \ref{Prop:1}}}
\label{Proof_P1}
In problem \eqref{WMD}, assuming $\mathbf{Q}(t+1)$ and $m^{\prime}(t+1)$ are given\footnote{In particular, the choice of $m^{\prime}(t+1)$ is independent with problem \eqref{WMD} and thus $m^{\prime}(t+1)$ can be selected from $[1,2,\cdots,M]$ following any given scheme in {Table \ref{table:WMD}}.}, the sufficient condition of the constraint in \eqref{WMD}
, i.e.,
\begin{align}
    & \mathbf{v}_{m^{\prime} (t+1) }^{H} \mathbf{Q}(t+1)^{H} \mathbf{\hat{C}}(t) \mathbf{Q}(t+1) \mathbf{v}_{m^{\prime}(t+1)} \nonumber\\
    & ~~~~\geq \mathbf{v}^H_{m_0(t+1)} \mathbf{Q}(t+1)^{H} \mathbf{\hat{C}}(t) \mathbf{Q}(t+1) \mathbf{v}_{m_0(t+1)}
\end{align}
is that the unit-norm vector $\mathbf{v}_{m^{\prime}(t+1)}$ is the first principal eigenvector of $\mathbf Q(t+1)^{H}\mathbf{\hat{C}}(t)\mathbf Q(t+1)$. This condition can be expressed as 
\begin{align}
 \mathbf{v}_{m^{\prime}(t+1)}  = \arg\max_{||\mathbf w || =1}  \mathbf w^H \mathbf Q(t+1)^{H}\mathbf{\hat{C}}(t)\mathbf Q(t+1) \mathbf w,\label{pro_p1_1}
\end{align}
which motivates us to acquire  $\mathbf Q(t+1)$ via two steps. 

\noindent 1).  \underline{\it Auxiliary matrix construction:}
we construct an auxiliary positive semi-definite (PSD) matrix $\mathbf R_{t+1}$ as follows
\begin{align}
   &\mathbf R_{t+1} \triangleq \sigma_{1} \mathbf{v}_{m^{\prime}(t+1)} \mathbf{v}_{m^{\prime}(t+1) }^{H}+\sum_{i=2}^{N_P}\sigma_{i}\mathbf{u}_{i-1} \mathbf{u}_{i-1}^{H}, \label{pro_p1_R_def}
\end{align}
where $\{\sigma_{n} \}_{n=1}^{N_P}$ and $\{\mathbf{u}_{n}\}_{n=1}^{N_P-1}$ are selected such that
\begin{align}
   &\sigma_{1}  \geq \sigma_{2}  \geq  \cdots \geq \sigma_{N_{P}}  >0,\\
   &\{ \mathbf{v}_{m^{\prime}(t+1)},\mathbf{u}_{1}  ,\cdots, \mathbf{u}_{N_{P}-1}  \} \text{ are all orthonormal.}
\end{align}
It is easy to check $\mathbf{v}_{m^{\prime}(t+1)}$ is the first principal eigenvector of the constructed $\mathbf{R}_{t+1}$, i.e., 
\begin{align}
	\mathbf{v}_{m^{\prime}(t+1)}  = \arg\max_{||\mathbf w || =1}  \mathbf w^H \mathbf R_{t+1} \mathbf w. \label{pro_p1_2}
\end{align}
\noindent 2).  \underline{\it Construct $\mathbf Q(t+1)$ :} In particular, $\mathbf Q(t+1)$ can be constructed by solving 
\begin{align}
	\mathbf Q(t+1)^{H}\mathbf{\hat{C}}(t)\mathbf Q(t+1) = \mathbf R_{t+1}. \label{pro_p1_6}
\end{align}
Notice that when $\mathrm{rank}(\mathbf{\hat C}(t)) = N_A>N_P$, there must exist a $\mathbf Q(t+1)$ such that
\begin{align}
    \mathrm{rank}(\mathbf Q(t+1)^{H}\mathbf{\hat{C}}(t)\mathbf Q(t+1)) = \mathrm{rank}(\mathbf R_{t+1}) = N_P,
\end{align}
which means the equation \eqref{pro_p1_6} is solvable.
Since equation \eqref{pro_p1_6} is a quadratic form of variable $\mathbf{Q}(t+1)$, it can be solved by completing the square. More specifically, due to the positive semi-definiteness of both $\mathbf{\hat{C}}(t)$ and $\mathbf R_{t+1}$, it can be shown 
\begin{align}
 \mathbf{\hat{C}}(t)  &=   \mathbf{\hat{C}}(t)^{\frac{1}{2}} \mathbf{\hat{C}}(t)^{\frac{1}{2}}, \label{pro_p1_7}\\
\mathbf R_{t\!+\!1} &=  \mathbf{Y}^H_{t\!+\!1} \mathbf \Sigma^H_{t\!+\!1} \mathbf{X}^H_{t\!+\!1} \mathbf X_{t\!+\!1} \mathbf \Sigma_{t\!+\!1} \mathbf Y_{t\!+\!1} , \label{pro_p1_8}
\end{align}
where 
\begin{align}
	\mathbf \Sigma_{t+1}  &= \mathrm{diag}\Big(\!\{  \sqrt{\sigma_{1}}, \sqrt{\sigma_{2}},\! \cdots\!,  \sqrt{\sigma_{N_P}} ,  0, \cdots, 0  \}\!\Big) \!\in \!\mathbb{C}^{N_A \!\times\! N_A}, \\
	 \mathbf Y_{t+1}   &= \left [ \mathbf{v}_{m^{\prime}(t+1)}, \mathbf{u}_{1}, \cdots,   \mathbf{u}_{N_P-1} \right] \!\in\! \mathbb{C}^{N_A \!\times\! N_P}, \label{pro_p1_Y}
\end{align} and $\mathbf X_{t+1} \in \mathbb{C}^{N_A \times N_A}$ is a random unitary matrix. 
From \eqref{pro_p1_7} and  \eqref{pro_p1_8}, equation \eqref{pro_p1_6} holds if 
\begin{align}
\mathbf{\hat{C}}(t)^{\frac{1}{2}}\mathbf Q(t+1) = \mathbf X_{t+1} \mathbf \Sigma_{t+1} \mathbf Y_{t+1},\label{36}
\end{align}
from which $\mathbf Q(t+1) $ can be obtained via 
\begin{align}
\mathbf Q(t+1) = \hat{\mathbf{C}}(t)^{-\frac{1}{2}} \mathbf X_{t+1} \mathbf \Sigma_{t+1}  \mathbf Y_{t+1}.
\end{align}
 
\subsection{The Proof of \textbf{Proposition \ref{Prop:2}}}
\label{Proof_P2}
In this proposition, the rank of $\mathbf{\hat{C}}(t)$ is $ K < N_A$ instead of $N_A$ which is assumed in \textbf{Proposition} \ref{Prop:1}
This subtle difference suggests that the proofs of these two propositions follow the same proof logic, in which the only difference is the construction of $\mathbf{R}_{t+1}$. 

Similarly, given $m^{\prime}(t+1)$, the auxiliary PSD matrix $\mathbf{R}_{t+1}$ is constructed as follows:
\begin{align}
    &\mathbf R_{t+1} = \sigma_{1} \mathbf{v}_{m^{\prime}(t+1)} \mathbf{v}_{m^{\prime}(t+1) }^{H}+\sum_{i=2}^{N_P}\sigma_{i}\mathbf{u}_{i-1} \mathbf{u}_{i-1}^{H} , 
\end{align}
where $\{\mathbf{u}_{n}\}_{n=1}^{N_P-1}$ are preselected such that columns $\{ \mathbf{v}_{m^{\prime}(t+1)},\mathbf{u}_{1} ,\cdots, \mathbf{u}_{N_P-1} \}$ are all orthonormal.
When $K < N_P$, the hyper-parameters $\{\sigma_{n}\}_{n=1}^{N_P}$ follow  
\begin{align}
        \sigma_{1} \geq \sigma_{2} \geq  \cdots \geq \sigma_{K} \textgreater \sigma_{K+1}  =  \cdots = \sigma_{N_P} = 0. \label{pro_p2_cond_1}
    \end{align}
Otherwise, when $N_P \leq K <N_A$, there is
\begin{align}
        \sigma_{1} \geq \sigma_{2} \geq  \cdots \geq  \sigma_{N_P} > 0.
\end{align}
In particular, $\mathbf Q(t+1)$ is obtained via solving 
\begin{align}
	\mathbf Q(t+1)^{H}\mathbf{\hat{C}}(t)\mathbf Q(t+1) = \mathbf R_{t+1}. \label{pro_p2_1}
\end{align} 
Therefore, we similarly consider the completing square form of equation \eqref{pro_p2_1}. Due to the positive semi-definiteness of  $\mathbf{\hat{C}}(t)$ and $\mathbf R_{t+1}$, there is
\begin{align}
	 \mathbf{\hat{C}}(t)^{\frac{1}{2}} \mathbf{\hat{C}}(t)^{\frac{1}{2}} &=  \mathbf{\hat{C}}(t), \label{pro_p2_2}\\
	 \mathbf{Y}^H_{t+1} \mathbf \Sigma^H_{t+1} \mathbf{X}^H_{t+1} \underbrace{\mathbf X_{t+1} \mathbf \Sigma_{t+1} \mathbf Y_{t+1}}_{\triangleq\mathbf F_{t+1}}  & = \mathbf R_{t+1}, \label{pro_p2_3}
\end{align}
where

\begin{align}
	&\mathbf Y_{t+1}   = \left [ \mathbf{v}_{m^{\prime}(t+1)}, \mathbf{u}_{1}, \cdots,   \mathbf{u}_{N_{P}-1} \right],\\
	&\mathbf \Sigma_{t+1} \!=
\!\!	\left[
	\begin{array}{c}
		\!\mathrm{diag}\left(\{ \!\sqrt{\sigma_{1}}, \sqrt{\sigma_{2}}, \cdots,  \sqrt{\sigma_{N_P}}\}\right) \! \in \!\mathbb{C}^{N_P\!\times\! N_P}\!\!\!\\ \hline 
		\mathbf{0} \in\mathbb{C}^{(N_A-N_P) \times N_P }
	\end{array}
	\right],\\
	&\mathbf X^H_{t+1} \mathbf X_{t+1} \!= \!\left[
	\begin{array}{c|c}
		\mathbf{I}_{K \times K}\! \in\mathbb{C}^{K\times K}& \mathbf 0 \in\mathbb{C}^{(N_A-K) \times K}\\ \hline 
		\mathbf 0 \in\mathbb{C}^{K \times (N_A-K)}& \mathbf 0 \in\mathbb{C}^{(N_A-K) \times (N_A-K)}
	\end{array}\!\!
	\right].
\end{align}
Then, from \eqref{pro_p2_1}, \eqref{pro_p2_2}, and \eqref{pro_p2_3},
the equation
\begin{align}
	\mathbf{\hat{C}}(t)^{\frac{1}{2}}\mathbf Q(t+1) = \mathbf F_{t+1} \label{pro_p2_4}
\end{align}
can be straightforwardly obtained. By replacing $\mathbf{\hat{C}}(t)^{\frac{1}{2}}$ in equation \eqref{pro_p2_4} with its eigenvalue decomposition
\begin{align}\label{pro_p2_5}
	\mathbf{\hat{C}}(t)^{\frac{1}{2}} = \left[
\begin{array}{c|c}
\mathbf{U}_{t,1} 
&\mathbf U_{t,2}
\end{array}
\right] 
\left[
\begin{array}{c|c}
\mathrm{diag}(\mathbf{s}_t) & \mathbf 0 \\ \hline 
\mathbf 0 & \mathbf 0 
\end{array}
\right]  
\left[
\begin{array}{c}
\mathbf U^{H}_{t,1}  \\ \hline 
\mathbf U^{H}_{t,2}
\end{array}
\right], 
\end{align} 
where $\mathbf{s}_t = \left[ \sigma^{c}_{t,1},  \cdots, \sigma^{c}_{t,K} \right]$ contains all nonzero singular values of $\mathbf{\hat{C}}(t)^{\frac{1}{2}}$ and the semi-unitary matrices $\mathbf U^{H}_{t,1}$ and $\mathbf U^{H}_{t,2}$ are orthogonal (i.e., $\mathbf U^{H}_{t,1}\mathbf U_{t,2} = \mathbf{0}$),
we have 
\begin{align}
&\left[
\begin{array}{c|c}
\mathbf U_{t,1} 
&\mathbf U_{t,2}
\end{array}
\right] 
\left[
\begin{array}{c|c}
\mathrm{diag}(\mathbf{s}_t) & \mathbf 0 \\ \hline 
\mathbf 0 & \mathbf 0 
\end{array}
\right]  
\left[
\begin{array}{c}
\mathbf U^{H}_{t,1}  \\ \hline 
\mathbf U^{H}_{t,2}
\end{array}
\right]\mathbf Q(t+1) =\mathbf F_{t+1}. \label{pro_p2_eq_1}
\end{align} 
Furthermore, due to $\mathbf U^{H}_{t,1}\mathbf U_{t,2} = \mathbf{0}$,  \eqref{pro_p2_eq_1} is equivalent to 
\begin{align}
&\left[
\begin{array}{c|c}
\mathrm{diag}(\mathbf{s}_t) & \mathbf 0 \\ \hline 
\mathbf 0 & \mathbf 0 
\end{array}
\right]  
\left[
\begin{array}{c}
\mathbf U^{H}_{t,1}  \\ \hline 
\mathbf U^{H}_{t,2}
\end{array}
\right] \!\mathbf Q(t\!+\!1) \!=\!\left[
\begin{array}{c}
	\mathbf U^{H}_{t,1}  \\ \hline 
	\mathbf U^{H}_{t,2}
\end{array}
\right] \mathbf F_{t+1}.
\end{align}
Following this, it holds that 
\begin{align}
\left[
\begin{array}{c}
\mathrm{diag}(\mathbf{s}_t)\mathbf U^{H}_{t,1} \mathbf Q(t+1) \\ \hline 
\mathbf{0}
\end{array}
\right]  = \left[
\begin{array}{c}
	\mathbf U^{H}_{t,1} \mathbf F_{t+1}   \\ \hline 
	\mathbf U^{H}_{t,2}\mathbf F_{t+1}  
\end{array}
\right] \label{pro_p2_6},
\end{align}
where
\begin{align}
&\mathbf U^{H}_{t,1} = \left[
\begin{array}{c|c}
\mathbf U^{H}_{t,11} \in \mathbb{C}^{K\times K} 
&\mathbf U^{H}_{t,12} \in \mathbb{C}^{K\times (N_A-K)}
\end{array}
\right], \\
&\mathbf Q(t+1) = \begin{bmatrix}
\begin{array}{c} 
  \mathbf{Q}_{1}(t+1) \in \mathbb{C}^{K\times K} \\ \hline 
   \mathbf{Q}_{2}(t+1) \in  \mathbb{C}^{(N_A-K)\times K} 
\end{array}
\end{bmatrix}.
\label{pro_p2_6_1}
\end{align}
From \eqref{pro_p2_6}, equation \eqref{pro_p2_4} has solutions if and only if 
\begin{align}
&\mathrm{diag}(\mathbf{s}_t)\left(\mathbf U^{H}_{t,11} \mathbf{Q}_{1}(t+1) + \mathbf U^{H}_{t,12} \mathbf{Q}_{2}(t+1)\right) =   \mathbf U^{H}_{t,1} \mathbf F_{t+1} ;  \label{pro_p2_7}\\
&\mathbf{0}= \mathbf U^{H}_{t,2} \mathbf F_{t+1}= \mathbf U^{H}_{t,2}\mathbf X_{t+1} \mathbf \Sigma_{t+1} \mathbf Y_{t+1}. \label{pro_p2_8}
\end{align}
To make equation \eqref{pro_p2_8} hold, we construct $\mathbf X_{t+1}$ as following: 
\begin{align}
\mathbf X_{t+1} = \mathbf{X}_{t+1, 1} \mathbf{X}_{t+1, 2},
\end{align}where 
\begin{align}
&\mathbf{X}_{t+1, 1}\!=\!
\left[
\begin{array}{c|c}
\mathbf{U}_{t,1}\in \mathbb{C}^{N_A \times K} & \mathbf{0} \in \mathbb{C}^{N_A \times (N_A - K)}
\end{array}
\right ] ,\label{pro_p2_X_decomp}\\
&\mathbf{X}_{t+1,2}\!= \!
\left[
\begin{array}{c|c}
\mathbf{V}_{t+1} \!\in\! \mathbb{C}^{K \!\times\! K} & \mathbf 0 \!\in\!\mathbb{C}^{(N_A\!-\!K) \!\times \!K}\\ \hline 
\mathbf 0 \!\in\!\mathbb{C}^{K \!\times\! (N_A\!-\!K)}& \mathbf 0 \!\in\!\mathbb{C}^{(N_A\!-\!K) \!\times\! (N_A\!-\!K)}
\end{array}
\right ],\\
&\mathbf{V}^H_{t+1} \mathbf{V}_{t+1} = \mathbf{I}.\label{pro_p2_X_3}
\end{align}
In particular, $\mathbf{X}_{t+1, 1}$ is designed carefully such that 
\begin{align}
    \mathbf U^{H}_{t,2}\mathbf X_{t+1} = \mathbf U^{H}_{t,2}\mathbf X_{t+1,1}\mathbf X_{t+1,2} = \mathbf{0}.
\end{align}
Then, using the specially designed $\mathbf X_{t+1}$, we can construct a special $\mathbf{Q}(t+1)$ to make equation \eqref{pro_p2_7} hold, that is, 
\begin{align}
\!\!\mathbf Q(t\!+\!1) \! = \!\!\left[
\!\begin{array}{c}
\mathbf{Q}_1(t\!+\!1) \\  
\mathbf{Q}_2(t\!+\!1)  
\end{array}\!
\right ]
\!\!=\!\! \left[
\!\!\begin{array}{c}
\mathbf{U}^{-H}_{t,11} 
\mathrm{diag}(\mathbf{s}_t)^{-1} \mathbf{U}^{H}_{t,1} \mathbf F_{t+1} \\ 
\mathbf{0}
\end{array}
\!\!\!\right ].\label{pro_p2_solu}
\end{align}
Finally, together with the null space of $\mathbf U^{H}_{t,1}$, we obtain the solution set of  \eqref{pro_p2_4} based on \eqref{pro_p2_solu} with the following form:
\begin{align}
\mathbf Q(t+1) = \left[
\begin{array}{c}
\mathbf{U}^{-H}_{t,11} \mathrm{diag}(\mathbf{s}_{t})^{-1} \mathbf{U}^{H}_{t,1} \mathbf F_{t+1} \\ 
\mathbf{0} 
\end{array}
\right ]+ \mathbf{O}_{t}, 
\end{align} where $\mathbf{O}_{t} \in  \mathrm{Null} (\mathbf{U}_{t,1}^H)$.

\vspace{-0.2cm}
\subsection{\!The \!Strategy \!1 \!of  $\mathbf{X}_{t+1}$\! in \textbf{Proposition \ref{Prop:1}} and \textbf{Proposition \ref{Prop:2}}}
\label{History_Q_info}
See Section \ref{subsec:3} of the supplementary document.
\vspace{-0.2cm}
\subsection{Strategy for Setting the Upper Bound $b$} 
\label{Strategy_b}
See Section \ref{subsec:4} of the supplementary document.
 \vspace{-0.2cm}
\subsection{Proof of \textbf{Proposition \ref{Prop:3}}} 
\label{Strategy_a}
See Section \ref{subsec:5} of the supplementary document.
\vspace{-0.2cm}
\subsection{The Proof of {\bf Property \ref{property:1}}}
\label{property1}
 See Section \ref{subsec:6} of the supplementary document.
\vspace{-0.2cm}
\subsection{The Proof of \textbf{Lemma \ref{Lemma:1}}}
\label{Covergence}
Regarding problem \eqref{relaxtion_problem}, its corresponding Lagrangian function is 
\begin{align}
&\sum_{i=1}^{t}\sum_{m=1}^M \frac{1}{\eta(i)} \log \Big ( \mathbf v_{m_0(i)}^H \mathbf Q(i)^H {\mathbf C} \mathbf Q(i) \mathbf v_{m_0(i)} \nonumber\\
&~~~~~~~~~~~~~~~~~~~~~~~~~~~~~- \mathbf v_m^H \mathbf Q(i)^H {\mathbf C} \mathbf Q(i) \mathbf v_m  \Big ) \nonumber\\
&~~~+ \log \det ( {\mathbf C} ) - \lambda ||{\mathbf C}||_{*} +\lambda^t_1 (\mathrm{Tr}({\mathbf C}) -b),
\end{align}
where $\lambda^t_1 \geq 0$ is the dual variable. Therefore, the optimal solution $\mathbf{\tilde C}_t$ of the problem \eqref{relaxtion_problem} satisfies the following KKT condition
\begin{small}
\begin{align}
    0=&\sum_{i=1}^{t}\sum_{m=1}^M\frac{\mathbf{Q}(i)\left(\mathbf{v}_{m_0(i)}\mathbf{v}^H_{m_0(i)}-\mathbf{v}_{m}\mathbf{v}^H_{m}\right)\mathbf{Q}^H(i)}{\eta(i)\!\!\left(\mathbf{v}^H_{\!m_0(i)\!}\mathbf{Q}^H\!(i)\mathbf{\tilde C}_t \mathbf{Q}(i)\mathbf{v}_{\!m_0(i)\!}\!-\!\mathbf{v}^H_{m}\mathbf{Q}^H(i)\mathbf{\tilde C}_t \mathbf{Q}(i)\mathbf{v}_{m}\right)} \nonumber\\
    & ~~+\mathbf{\tilde C}^{-H}_t-\lambda\mathbf{I} + \lambda^{*,t}_1\mathbf{I}; \label{conv_1}\\
    =& \sum_{i=1}^t \sum_{m=1}^M \frac{\mathbf{G}_{m,i}}{\eta(i)\langle\mathbf{G}_{m,i},\mathbf{\tilde C}_t\rangle_F} + \mathbf{\tilde C}^{-H}_t-\lambda\mathbf{I} + \lambda^{*,}t_1\mathbf{I} \label{conv_2},
\end{align}
\end{small}where $\lambda^{*,t}_1$ represent the optimal dual variable.
Based on \textbf{Definition \ref{def:notations_2}}, by multiplying $\mathbf{C} \in \{\mathbf{C}| \mathrm{Tr}({\mathbf C} ) \leq b\}$ to both sides of equation \eqref{conv_2}, there is 
\begin{align}
    &\sum_{i=1}^t \sum_{m=1}^M \frac{\langle\mathbf{G}_{m,i},\mathbf{C}\rangle_F}{\eta(i)\langle\mathbf{G}_{m,i},\mathbf{\tilde C}_t\rangle} +\langle\mathbf{\tilde C}^{-H}_t,\mathbf{C}\rangle_F + (\lambda^{*,t}_1- \lambda)\langle\mathbf{I},\mathbf{C}\rangle_F,
    \label{conv_3}
\end{align}
which is equal to
\begin{align}
       S_t\mu_t(\mathbf{C}) + \langle\mathbf{\tilde C}^{-H}_t,\mathbf{C}\rangle_F + (\lambda^{*,t}_1-\lambda)\langle\mathbf{I},\mathbf{C}\rangle_F&= 0.
    \label{conv_4}
\end{align}
Since $\mathbf{C}, \mathbf{\tilde C}_t$ are positive semi-definite matrix and $\lambda^{*,t}_1\geq0$, the term $\langle\mathbf{\tilde C}^{-H}_t,\mathbf{C}\rangle_F$ and $\lambda^{*,t}_1\langle\mathbf{I},\mathbf{C}\rangle_F$ in equation \eqref{conv_4} are positive such that 
\begin{align}
     S_t\mu_t(\mathbf{C}) \leq \lambda \langle\mathbf{I},\mathbf{C}\rangle_F.
\end{align}
According to the definition of $S_t$, the $S_t$ is positive such that
\begin{align}
    \mu_t(\mathbf{C}) \leq \frac{\lambda \langle\mathbf{I},\mathbf{C}\rangle_F}{S_t}
\end{align}
holds for all $\mathbf{C} \in \{\mathbf{C}| \mathrm{Tr}({\mathbf C} ) \leq b\}$. 
 
{\color{black} In particular, as shown in {\bf Theorem \ref{theorem:divergent_sequence}}, $\{S_t\}_t$ is a divergent sequence under the following assumption. 
\vspace{-0.15cm}
\begin{assumption}
\label{assumption:1}
  For any $t$, there exists a real number $l$ such that $\lambda_{min}(\mathbf{\tilde{C}}_t) \geq l$, where $\lambda_{min}(\mathbf{\tilde{C}}_t)$ is the smallest non-zero eigenvalue of $\mathbf{\tilde{C}}_t$.
\end{assumption}
\vspace{-0.3cm}
\begin{theorem}
\label{theorem:divergent_sequence}
   For any real number $N>0$, there always exists a natural number $T$ such that $S_t\geq N$ holds for all $t>T$. 
  \begin{proof}

  According to {\bf Definition \ref{def:notations_2}}, the term $\eta(i)\langle\mathbf{G}_{m,i},\mathbf{\tilde C}_t\rangle_F$ is the reciprocal of summand of $S_t$ (i.e., $\omega_{i,m,t}$), and such a term can be upper bounded as 
  \begin{align}
      &\eta(i) \langle\mathbf{G}_{m,i},\mathbf{\tilde C}_t\rangle_F \nonumber\\
      &~~~= \left(\mathbf{v}_{m_0(i)}^H\mathbf{Q}^H(i){\tilde{\mathbf{C}}}_t\mathbf{Q}(i)\mathbf{v}_{m_0(i)}\right) \times\nonumber\\
      &~~~~~~~~~\left(\mathbf{v}^H_{m_0(i)}\mathbf{Q}^H(i)\mathbf{\tilde C}_t \mathbf{Q}(i)\mathbf{v}_{m_0(i)}-\mathbf{v}^H_m\mathbf{Q}^H(i) \mathbf{\tilde C}_t \mathbf{Q}(i)\mathbf{v}_m\right) \nonumber\\
      &~~~\leq \left(\mathbf{v}_{m_0(i)}^H\mathbf{Q}^H(i){\tilde{\mathbf{C}}}_t\mathbf{Q}(i)\mathbf{v}_{m_0(i)}\right)^2 \nonumber \\
       &~~~\leq \left(\mathrm{Tr}\left(\mathbf{v}_{m_0(i)}\mathbf{v}_{m_0(i)}^H\right)\mathrm{Tr}\left(\mathbf{Q}^H(i){\tilde{\mathbf{C}}}_t\mathbf{Q}(i)\right)\right)^2,\label{ineq1}
    \end{align}
    where $\mathbf{v}_{m_0(i)}$ is unitary, i.e., $\mathrm{Tr}\left(\mathbf{v}_{m_0(i)}\mathbf{v}_{m_0(i)}^H\right) = 1$.
   Notice that $\mathbf{Q}(i)$ is formulated as
    \begin{align}
      \mathbf Q(i) = 
      \underbrace{\left[
      \begin{array}{c}
      \mathbf{U}^{-H}_{i-1,11} \mathrm{diag}(\mathbf{s}_{i-1})^{-1} \mathbf{U}^{H}_{i-1,1}\mathbf{F}_{i}\\ 
      \mathbf{0} 
      \end{array}
      \right ]}_{\mathbf{\tilde Q}(i)} + \mathbf{O}_{i-1},  \label{Q_form}
      \end{align}
    where orthogonal matrix $\mathbf{O}_{i-1}$ can be chosen such that ${\tilde{\mathbf{C}}}_t\mathbf{O}_{i-1} = \mathbf{0}$ (see \eqref{O_def}). 
    Substituting \eqref{Q_form} into \eqref{ineq1}, \eqref{ineq1} can further be expressed as 
 \begin{align}
      \eta(i) \langle\mathbf{G}_{m,i},\mathbf{\tilde C}_t\rangle_F &\leq \left(\mathrm{Tr}\left(\mathbf{\tilde Q}^H(i){\tilde{\mathbf{C}}}_t\mathbf{\tilde Q}(i)\right)\right)^2 \nonumber\\
      &\leq \left(\mathrm{Tr}\left({\tilde{\mathbf{C}}}_t\right)\mathrm{Tr}\left(\mathbf{\tilde Q}^H(i)\mathbf{\tilde Q}(i)\right)\right)^2.
     \label{ineq2_1}
    \end{align}
    Moreover, using trace inequality and $\mathrm{Tr}\left({\tilde{\mathbf{C}}}_t\right) \leq b$, \eqref{ineq2_1} can be relaxed as 
    \begin{align}
         &\eta(i) \langle\mathbf{G}_{m,i},\mathbf{\tilde C}_t\rangle_F \nonumber\\
         &~~~\leq  \Big(b\mathrm{Tr}\left(\mathbf{F}^H_{i}\mathbf{F}_{i} \right) \mathrm{Tr}\left( \mathbf{U}_{i-1,1} \mathbf{U}^H_{i-1,1}\right) \times \nonumber\\
      &~~~~~~~~~~~\mathrm{Tr} \left(\mathbf{U}^{-H}_{i-1,11} \mathrm{diag}(\mathbf{s}^2_{i-1})^{-1} \mathbf{U}^{-1}_{i-1,11}  \right)\Big)^2. \label{ineq2}
    \end{align}
     Based on \eqref{notation_F}, \eqref{notation_U} and \eqref{notation_diag_s}, it is straightforward to check 
    \begin{align}
      \mathrm{Tr}\left(\mathbf{F}^H_{i}\mathbf{F}_{i} \right) &\leq N_A; \label{tr1}\\
      \mathrm{Tr}\left( \mathbf{U}_{i-1,1} \mathbf{U}^H_{i-1,1}\right) & \leq K <N_A; \label{tr2}\\
      \mathbf{U}^{-H}_{\!i\!-\!1,11} \mathrm{diag}(\mathbf{s}^2_{i\!-\!1})^{-1} \mathbf{U}^{-1}_{\!i\!-\!1,11} &= \Big(\underbrace{\mathbf{U}^{H}_{\!i\!-\!1,11} \mathrm{diag}(\mathbf{s}^2_{i\!-\!1}) \mathbf{U}_{\!i\!-\!1,11}}_{{\tilde{\mathbf{C}}}_{i|K}}\Big)^{-1},
    \end{align}
    where ${\tilde{\mathbf{C}}}_{i|K}$ is the $K$-th order leading principal submatrix of ${\tilde{\mathbf{C}}}_{i}$. 
    Using Theorem 2.1 in \cite{HORN199829} and {\bf Assumption \ref{assumption:1}}, the relationship between the smallest eigenvalue of ${\tilde{\mathbf{C}}}_{i}$ and of ${\tilde{\mathbf{C}}}_{i|K}$ can be specified as
    \begin{align}
      l \leq \lambda_{min}({\tilde{\mathbf{C}}}_{i}) \leq \lambda_{min}({\tilde{\mathbf{C}}}_{i|K}). \label{ineq3}
    \end{align}
    Furthermore, since the eigenvalues of ${\tilde{\mathbf{C}}}_{i|K}^{-1}$ are exactly reciprocal of the eigenvalues of ${\tilde{\mathbf{C}}}_{i|K}$, the upper bound of $\mathrm{Tr}({\tilde{\mathbf{C}}}_{i|K}^{-1})$ can be deduced as follows
    \begin{align}
      \mathrm{Tr}({\tilde{\mathbf{C}}}_{i|K}^{-1}) \leq \frac{K}{\lambda_{min}({\tilde{\mathbf{C}}}_{i|K} )}  \leq \frac{N_A}{l}. \label{tr3}
    \end{align}
    By substituting \eqref{tr1}, \eqref{tr2} and \eqref{tr3} into \eqref{ineq2}, the upper bound can be obtained
    \begin{align}
      \eta(i) \langle\mathbf{G}_{m,i},\mathbf{\tilde C}_t\rangle_F \leq \frac{b^2 N^6_A}{l^2}.
    \end{align} 
    
    By letting $T = \lceil\frac{Nb^2N_A^6}{Ml^2}\rceil$, it is easy to show that 
    \begin{align}
        S_t = \sum_{i=1}^{t}\sum_{m=1}^M\frac{1}{\eta(i)\langle \mathbf{G}_{m,i},\mathbf{\tilde C}_t\rangle_F} \geq \frac{tMl^2}{b^2N_A^6} \geq N.\end{align}
  \end{proof}
  \vspace{-0.3cm}
\end{theorem}
Due to the divergence of sequence $\{S_t\}_t$, $S_t$ tends to $\infty$ when $t$ goes to infinity, which results in the upper bound of $\mu_t(\mathbf{C})$ tends to $0$.}
\vspace{-0.3cm}
\subsection{The Proof of {\bf Lemma \ref{Lemma:2}}}
\label{Convergence_2}


According to {\bf Proposition \ref{Prop:1}} and {\bf Proposition \ref{Prop:2}}, for the communication round $\tau > t$, the proposed algorithm obtains an interior feasible CCM $\mathbf{\hat C}_{\tau} \in \mathcal{\bar F}_{\tau}$ to generate $\mathbf{Q}(\tau)$ . Based on such $\mathbf{Q}(\tau)$, the cutting planes passing through $\mathbf{\hat C}_{\tau}$ are constructed to cut $\mathcal{\bar F}_{\tau}$ (see Section \ref{sec:CCM Reconstruction}). 

\textcolor{black}{Based on the design rule of weighting matrix $\mathbf{Q}$ (see Section \ref{sec:WMD}), it is easy to show that the generated cutting planes can cut $\mathcal{\bar F}_{\tau}$, such that part of $\mathcal{\bar F}_{\tau}$ are excluded from $\mathcal{\bar F}_{\tau}$. Suppose that there are some feasible points, except for the ground-truth CCM (see {\bf Property \ref{property:2}} (iii)), can not be removed at any communication round. For any such feasible point, customized cutting planes can be constructed to remove this point (at least). Hence, the above-mentioned assumption can not hold, which further shows that as $\tau$ increases, $\mathcal{\bar F}_{\tau}$ will converge to the set that only contains $\mathbf{C}^*$.}


\vspace{-0.3cm}
\subsection{The Proof of \textbf{Theorem \ref{theorem:1}}}
\label{convergence_theorem}
Since it holds that the ground-truth CCM $\mathbf{C}^* \in \mathcal{\bar F}_t$ (see \textbf{Property \ref{property:2}}), the value of $\langle \mathbf{G}_{m,i}, \mathbf{C}^*\rangle_F, \forall m,i$ is always non-negative according to the definition of $\mathbf{G}_{m,i}$, which gives $\mu_t(\mathbf{C}^*) \geq 0$. Then, from \textbf{Lemma \ref{Lemma:1}}, the upper bound of $\mu_t(\mathbf{C}^*)$ tends to $0$ when $t$ is large enough due to $\mathbf{C}^* \in \{\mathbf{C}| \mathrm{Tr}({\mathbf C} ) \leq b\}$, thus it holds that 
\begin{align}
    0 \leq \mu_t(\mathbf{C}^*) \leq 0. 
\end{align}

For such $\mathcal{\bar F}_t$ with zeros size measure,  \textbf{Lemma \ref{Lemma:2}} guarantees it shrinks to a singleton. In other words, such $\mathcal{\bar F}_t$ only contains one element, that is $\mathbf{C}^*$. 
Based on the relationship $\mathcal{F}_t \subseteq \mathcal{\hat F}_t \subseteq \mathcal{\bar F}_t$ and $\mathbf{C}^* \in \mathcal{F}_t$ (from \textbf{Property \ref{property:1}}), it can be concluded that $\mathcal{F}_t$ also only contains one feasible CCM $\mathbf{C}^*$.
\vspace{-0.3cm}
\subsection{Numerical Results of Trying Different Strategies in Table \ref{table:WMD}} 
\label{strategies_table}
See Section \ref{subsec:7} of the supplementary document.

\bibliography{refs}
\bibliographystyle{IEEEtran}

\newpage

\noindent {\bf Supplementary Document for ``Downlink Channel Covariance Matrix Reconstruction For Limited Feedback FDD Massive MIMO Systems''.}
{
\setcounter{subsection}{0}
 \subsection{The Convex Cone Property of \eqref{notation_P}}
    \label{subsec:1}
    According to the definition of convex cone \cite{boyd}, it can be concluded that $\bigcap_{t=1}^T \mathcal{P}_t$ is a convex cone if and only if each $\mathcal{P}_t$ is a convex cone. 
    
    Given $\mathbf{\hat C} \in \mathcal{P}_t$ and $\theta\geq 0$, the inequalities  
    \begin{align}
        &\mathbf {v}^H_{m} \mathbf Q(i)^H \hat{\mathbf C} \mathbf Q(i) \mathbf v_{m} \leq  \mathbf v^H_{m_0(i)} \mathbf Q(i)^H \hat{\mathbf C} \mathbf Q(i) \mathbf v_{m_0(i)}, \nonumber\\
        &~~~~~~~~~~~~~~~~~~~~~~~~~~~~~~~~~~~i = 1,\ldots, t,~~ \forall \mathbf v_m \in \mathcal V \label{inq1}
    \end{align}
    is equivalent to
    \begin{align}
        &\theta \mathbf {v}^H_{m} \mathbf Q(i)^H \hat{\mathbf C} \mathbf Q(i) \mathbf v_{m} \leq  \theta  \mathbf v^H_{m_0(i)} \mathbf Q(i)^H \hat{\mathbf C} \mathbf Q(i) \mathbf v_{m_0(i)},\nonumber\\
        &~~~~~~~~~~~~~~~~~~~~~~~~~~~~~~~~~~~~i = 1,\ldots, t, \forall \mathbf v_m \in \mathcal V
    \end{align} 
    by multiplying $\theta$ on both sides of \eqref{inq1}.
    Hence, it can be concluded that $\theta \mathbf{\hat C} \in \mathcal{P}_t$, which means $\mathcal{P}_t$ is a cone. 
    
    In addition, given $\mathbf{\hat C}_1, \mathbf{\hat C}_2 \in \mathcal{P}_t$ and $\theta \in [0,1]$, 
    it holds that 
    \begin{align}
        &~\mathbf {v}^H_{m} \mathbf Q(i)^H \left(\theta \mathbf{\hat C}_1+ (1-\theta)\mathbf{\hat C}_2\right) \mathbf Q(i) \mathbf v_{m} \nonumber\\
        =&~\mathbf {v}^H_{m} \mathbf Q(i)^H \left(\theta \mathbf{\hat C}_1\right) \mathbf Q(i) \mathbf v_{m} \!+\! \mathbf {v}^H_{m} \mathbf Q(i)^H \left((1-\theta) \mathbf{\hat C}_2\right) \mathbf Q(i) \mathbf v_{m} \nonumber\\
        \leq&~\mathbf v^H_{m_0(i)} \mathbf Q(i)^H \left(\theta \mathbf{\hat C}_1\right) \mathbf Q(i) \mathbf v_{m_0(i)} \nonumber\\
        & ~~~~~~+ \mathbf v^H_{m_0(i)} \mathbf Q(i)^H \left((1-\theta) \mathbf{\hat C}_2\right) \mathbf Q(i) \mathbf v_{m_0(i)} \nonumber\\
        =&~\mathbf v^H_{m_0(i)} \mathbf Q(i)^H \left(\theta \mathbf{\hat C}_1 + (1-\theta) \mathbf{\hat C}_2\right) \mathbf Q(i) \mathbf v_{m_0(i)}, \nonumber\\
        &~~~~~~~~~~~~~~~~~~~~~~~~~~~~~~~~~~~~~~i = 1,\ldots, t,~~ \forall \mathbf v_m \in \mathcal V. \label{inq2}
    \end{align}
    In other words, from \eqref{inq2}, we have $\theta \mathbf{\hat C}_1+ (1-\theta)\mathbf{\hat C}_2 \in \mathcal{P}_t$ which means $\mathcal{P}_t$ is convex. 
    
    Therefore, each $\mathcal{P}_t$ is a convex cone, and the union of these cones, i.e., $\bigcap_{t=1}^T \mathcal{P}_t$, is also a convex cone.

\subsection{The Shrinkage of $\mathcal B ( \mathbf{\hat C}; \{\mathbf Q(t)\}_{t=1}^T)$}
\label{subsec:2}
According to the definition of subset, it holds that 
\begin{align}
    \mathcal B ( \mathbf{\hat C}; \{\mathbf Q(t)\}_{t=1}^{\kappa_2}) \subseteq  \mathcal B ( \mathbf{\hat C}; \{\mathbf Q(t)\}_{t=1}^{\kappa_1}), ~~  \kappa_1 \leq \kappa_2 ,
\end{align} if and only if any $\hat{\mathbf{C}} \in \mathcal B ( \mathbf{\hat C}; \{\mathbf Q(t)\}_{t=1}^{\kappa_2})$ must be in $\mathcal B ( \mathbf{\hat C}; \{\mathbf Q(t)\}_{t=1}^{\kappa_1})$. From the definition of $\mathcal B ( \mathbf{\hat C}; \{\mathbf Q(t)\}_{t=1}^{T})$, it is easy to find that $\mathcal B ( \mathbf{\hat C}; \{\mathbf Q(t)\}_{t=1}^{\kappa_2})$ extends from $\mathcal B ( \mathbf{\hat C}; \{\mathbf Q(t)\}_{t=1}^{\kappa_1})$ through incorporating more constraints. Therefore, any $\mathbf{\hat C}$ in $\mathcal B ( \mathbf{\hat C}; \{\mathbf Q(t)\}_{t=1}^{\kappa_2})$ is also in $\mathcal B ( \mathbf{\hat C}; \{\mathbf Q(t)\}_{t=1}^{\kappa_1})$.

\subsection{\!The \!Strategy \!1 \!of  $\mathbf{X}_{t+1}$\! in \textbf{Proposition \ref{Prop:1}} and \textbf{Proposition \ref{Prop:2}}}
\label{subsec:3}
In this strategy, the history information $\{\mathbf{Q}(i)\}_{i=1}^{t}$ is used to design $\mathbf{Q}(t+1)$. The key idea is to require that $\mathbf{Q}(t+1)$ and $\{\mathbf{Q}(i)\}_{i=1}^{t}$ are as different as possible. Based on this idea,  the problem is formulated as follows
\begin{align}
    \min_{ \mathbf{Q}(t+1)} \operatorname{Tr}\left(\mathbf{Q}^{H}(t+1) \mathbf{B}(t)+\mathbf{B}^{H}(t) \mathbf{Q}(t+1)\right), \label{Q_info_1}
\end{align}
where $\mathbf{B}(t)=\sum_{i=1}^{t} \mathbf{Q}(i) \exp\{10(i-t)\}$. In particular, since the construction of $\mathbf{Q}(t+1)$ involves $\mathbf{X}_{t+1}$ which is a random matrix, this strategy is to determine $\mathbf{X}_{t+1}$ in essence. According to different proposition, the matrix $\mathbf{X}_{t+1}$ is computed as follows.

\noindent\subsubsection{\textbf{Proposition \ref{Prop:1}}} 
By replacing $\mathbf{Q}(t+1)$ with \eqref{p1_1}, the problem \eqref{Q_info_1} can be reformulated as 
\begin{align}
     &\min_{\mathbf{X}_{t+1}} \operatorname{Tr}\Big (\mathbf Y^H_{t+1} \mathbf \Sigma^H_{t+1} \mathbf X^H_{t+1} \hat{\mathbf{C}}(t)^{-\frac{1}{2}}\mathbf{B}(t) \nonumber\\
     &~~~~~~~~~~~~~~+\mathbf{B}^{H}(t) \hat{\mathbf{C}}(t)^{-\frac{1}{2}} \mathbf X_{t+1} \mathbf \Sigma_{t+1}  \mathbf Y_{t+1} \Big)\nonumber\\
    &~~~~~~~~\mathrm{s.t.} ~~~~\mathbf{X}^H_{t+1}\mathbf{X}_{t+1} = \mathbf{I}.
    \label{Q_info_2}
\end{align}
Based on the property of trace operation, the problem \eqref{Q_info_2} can be reformulated as 
\begin{align}
     &\max_{\mathbf{X}_{t+1}} \operatorname{Tr}\left(\mathbf X^H_{t+1} \mathbf{D}(t+1)+\mathbf{D}^{H}(t+1) \mathbf X_{t+1} \right)\nonumber\\
    &~~~~~~~~\mathrm{s.t.} ~~~~\mathbf{X}^H_{t+1}\mathbf{X}_{t+1} = \mathbf{I},
    \label{Q_info_3}
\end{align}
where 
\begin{align}
    \mathbf{D}(t+1) = - \hat{\mathbf{C}}(t)^{-\frac{1}{2}} \mathbf{B}(t) \mathbf Y^H_{t+1} \mathbf \Sigma^H_{t+1}. 
\end{align}
Then, from \cite{stiefel_manifold}, the closed-form solution of \eqref{Q_info_3} is 
\begin{align}
    \mathbf{X}^*_{t+1} = \tilde{\mathbf{U}}(t+1)\tilde{\mathbf{V}}^H(t+1),
\end{align}
where $\tilde{\mathbf{U}}(t+1)$ and $\tilde{\mathbf{V}}(t+1)$ are the left-orthonormal matrix and right-orthonormal matrix of $\mathbf{D}(t+1)$ respectively, which are from the singular value decomposition of $\mathbf{D}(t+1)$, i.e., $\mathbf{D}(t+1) = \tilde{\mathbf{U}}(t+1)\tilde{\mathbf \Phi}(t+1)\tilde{\mathbf{V}}^H(t+1)$.

\noindent\subsubsection{\textbf{Proposition \ref{Prop:2}}} 
Similarly, by replacing $\mathbf{Q}(t+1)$ with solution \eqref{p2_1}, the problem \eqref{Q_info_1} can be reformulated as 
\begin{align}
     &\max_{\mathbf{X}_{t+1}} ~~~\operatorname{Tr}\left(\mathbf X^H_{t+1} \mathbf{D}(t+1)+\mathbf{D}^{H}(t+1) \mathbf X_{t+1} \right)\nonumber\\
    &\mathrm{s.t.} ~~\mathbf{X}^H_{t+1}\mathbf{X}_{t+1} =  \begin{bmatrix}
  \mathbf{I}\in \mathbb{C}^{K\!\times \!K}  & \mathbf{0} \in \mathbb{C}^{K\!\times\! (N_A \!- \!K)}\\ 
  \mathbf{0} \in \mathbb{C}^{(N_A \!-\! K)\!\times\! K} & \mathbf{0} \in \mathbb{C}^{(N_A \!-\! K)\!\times\! (N_A \!-\! K)}
\end{bmatrix},
    \label{Q_info_4}
\end{align}
where 
\begin{align}
    \mathbf{D}(t+1) = - \left[
	\begin{array}{c}
	\mathbf{U}^{-H}_{t,11} \mathrm{diag}(\mathbf{s}_t)^{-1} \mathbf{U}^{H}_{t,1}\\ 
	\mathbf{0} 
	\end{array}
	\right ]^H \mathbf{B}(t) \mathbf Y^H_{t+1} \mathbf \Sigma^H_{t+1}. 
\end{align}
In particular, according to \eqref{pro_p2_X_decomp}-\eqref{pro_p2_X_3}, $\mathbf{X}_{t+1}$ is decided by $\mathbf{V}_{t+1}$. Therefore, the problem \eqref{Q_info_4} is further reformulated as 
\begin{align}
     &\max_{\mathbf{V}_{t+1}} \operatorname{Tr}\left(\mathbf V^H_{t+1} \tilde{\mathbf{D}}(t+1)+\tilde{\mathbf{D}}^{H}(t+1) \mathbf V_{t+1} \right)\nonumber\\
    &~~~~~~~~\mathrm{s.t.} ~~~~\mathbf{V}^H_{t+1}\mathbf{V}_{t+1} = \mathbf{I},
    \label{Q_info_5}
\end{align}
where 
\begin{align}
    \tilde{\mathbf{D}}(t+1) =  -\mathrm{diag}(\mathbf{s}_t)^{-1} \mathbf{U}_{t,11}^{-1} [\mathbf{B}(t) \mathbf Y^H_{t+1} \mathbf \Sigma^H_{t+1}]_{(1:K,1:K)},
\end{align}
and $ [\mathbf{B}(t) \mathbf Y^H_{t+1} \mathbf \Sigma^H_{t+1}]_{(1:K,1:K)}$ is the $k$-th leading principle submatrix of matrix $\mathbf{B}(t) \mathbf Y^H_{t+1} \mathbf \Sigma^H_{t+1}$.

Then, from \cite{stiefel_manifold}, the closed-from solution of \eqref{Q_info_5} is 
\begin{align}
    \mathbf{V}^*_{t+1} = \tilde{\mathbf{U}}(t+1)\tilde{\mathbf{W}}^H(t+1),
\end{align}
where $\tilde{\mathbf{U}}(t+1)$ and $\tilde{\mathbf{W}}(t+1)$ are the left-orthonormal matrix and right-orthonormal matrix of $\tilde{\mathbf{D}}(t+1)$ respectively, which are from the singular value decomposition of $\tilde{\mathbf{D}}(t+1)$, i.e., $\tilde{\mathbf{D}}(t+1) = \tilde{\mathbf{U}}(t+1)\tilde{\Phi}(t+1)\tilde{\mathbf{W}}^H(t+1)$. 
Using  $\mathbf{V}^*_{t+1}$, $\mathbf{X}^*_{t+1}$ can be computed directly from \eqref{pro_p2_X_decomp}-\eqref{pro_p2_X_3}.

\subsection{Strategy for Setting the Upper Bound $b$}
\label{subsec:4}
\textcolor{black}{Specifically, given the weight matrix $\mathbf{Q}$ and any codeword $\mathbf{v}$ in codebook $\mathcal V$, the beam precision of $\mathbf{Q}\mathbf{v}$ with respect to the ground-truth channel covariance $\mathbf{C}$ is defined as 
\begin{align}
    \frac{\mathbf{v}^H\mathbf{Q}^H\mathbf{C}\mathbf{Q}\mathbf{v}}{d \|\mathbf{Q}\mathbf{v}\|^2}, \label{strategy_b_beam_precision}
\end{align}
where $d$ is the largest eigenvalue of $\mathbf C$.  Due to the unknown of  $\mathbf{C}$, the value of beam precision of $\mathbf{Q}(t) \mathbf{v}_{m_0{(t)}}$ is difficult to obtain for the BS. However, the BS can utilize the historical CCM $\mathbf{C}^h$ (which is assumed to be known) to approximate this beam precision. By randomly generating numerous weight matrices $\{\mathbf{Q}_l\}_{l=1}^L$ , the BS can firstly use $\mathbf{C}^h$ to find a codeword set $\{\mathbf{v}_l\}_{l=1}^L$, where 
\begin{align}
    \mathbf{v}_l = \arg\max_{\mathbf{v}\in\mathcal{V}} \mathbf{v}^H\mathbf{Q}^H_l\mathbf{C}^h\mathbf{Q}_l\mathbf{v}.
\end{align} 
Then, by using pairs $\{(\mathbf{Q}_l,\mathbf{v}_l)\}_{l=1}^L$ and  $\mathbf{C}^h$, the beam precision of $\mathbf{Q}(t) \mathbf{v}_{m_0{(t)}}$ can be approximated as follows
\begin{align}
    \frac{\mathbf{v}^H_{m_0(t)} \mathbf{Q}(t)^H \mathbf{C} \mathbf{Q}(t) \mathbf{v}_{m_0(t)}} {d \|\mathbf{Q}(t) \mathbf{v}_{m_0{(t)}} \|^2} \approx \sum_{l} \frac{\mathbf{v}^H_{l} \mathbf{Q}_l^H \mathbf{C}^h \mathbf{Q}_l \mathbf{v}_{l}} {L d^h \|\mathbf{Q}_l \mathbf{v}_{l} \|^2}   \triangleq \alpha,
\end{align}
where $d^h$ is the largest eigenvalue of $\mathbf{C}^h$. Furthermore, using the definition of CQI, i.e., $\mathbf{v}^H_{m_0(t)} \mathbf{Q}(t)^H \mathbf{C} \mathbf{Q}(t) \mathbf{v}_{m_0(t)} = \eta(t) $, $d$ can be estimated as
\begin{align}
  & d \approx \frac{\eta(t)}{ \alpha \|\mathbf{Q}(t) \mathbf{v}_{m_0{(t)}} \|^2}.
\end{align}
Since $\mathrm{rank}(\mathbf{C}) = N_U$, we have $\mathrm{Tr}(\mathbf{C}) \leq N_U d$. Then, the upper bound can be obtained by 
\begin{align}
b = N_U d \approx \frac{N_U \eta(t) }{ \alpha \|\mathbf{Q}(t) \mathbf{v}_{m_0{(t)}} \|^2}. 
\end{align}
}
\subsection{Proof of \textbf{Proposition \ref{Prop:3}}} 
\label{subsec:5}
According to the definition of CQI (see \eqref{CQI}), in each communication round $t$, we have 
\begin{align}
\eta(t) &=  \mathbf{v}^{H}_{m_0(t)}\mathbf{Q}^H(t)\mathbf{C}\mathbf{Q}(t)\mathbf{v}_{m_0(t)} \nonumber \\
& = \mathrm{Tr}\left( \mathbf{v}^{H}_{m_0(t)}\mathbf{Q}^H(t)\mathbf{C}\mathbf{Q}(t)\mathbf{v}_{m_0(t)} \right)\nonumber \\
& = \mathrm{Tr}\left( \mathbf{C}\mathbf{Q}(t)\mathbf{v}_{m_0(t)} \mathbf{v}^{H}_{m_0(t)}\mathbf{Q}^H(t)\right). \label{pro_p3_1}
\end{align}
Notice that $\mathbf{A}_t \triangleq \mathbf{Q}(t)\mathbf{v}_{m_0(t)} \mathbf{v}^{H}_{m_0(t)}\mathbf{Q}^H(t)$ is a positive semi-definite matrix, Furthermore, $\mathbf{C}$ and $\mathbf{A}_t$ have the corresponding positive semi-definite square root matrix $\mathbf{C}^{\frac{1}{2}}$ and $\mathbf{A}_t^{\frac{1}{2}}$ respectively.
Therefore, from equation \eqref{pro_p3_1}, we have 
\begin{align}
    \eta(t) &= \mathrm{Tr}(\mathbf{C}\mathbf{A}_t)\nonumber\\
    & = \mathrm{Tr}(\mathbf{C}^\frac{1}{2}\mathbf{A}_t\mathbf{C}^\frac{1}{2})\nonumber\\
    & = \mathrm{Tr}(\mathbf{C}^\frac{1}{2}\mathbf{A}_t^\frac{1}{2}\mathbf{A}_t^\frac{1}{2}\mathbf{C}^\frac{1}{2})\nonumber\\
    & = \|\mathbf{C}^\frac{1}{2}\mathbf{A}_t^\frac{1}{2}\|^2_F \nonumber \\
    & \leq \|\mathbf{C}^\frac{1}{2}\|^2_F\|\mathbf{A}_t^\frac{1}{2}\|^2_F \nonumber \\
    & = \mathrm{Tr}(\mathbf C) \mathrm{Tr}(\mathbf A_t). \label{pro_p3_2}
\end{align}
Following this inequality \eqref{pro_p3_2}, there is 
\begin{align}
\mathrm{Tr}(\mathbf C) \ge \frac{\eta(t)}{\mathrm{Tr}\left(\mathbf{A}_t\right)} = \frac{\eta(t)}{\mathrm{Tr}\left(\mathbf{Q}(t) \mathbf{v}^{}_{m_0(t)} \mathbf{v}_{m_0(t)}^{H} \mathbf{Q}^{H}(t)\right)}.\label{pro_p3_3}
\end{align}
Since the inequality \eqref{pro_p3_3} holds for $\forall t \in \{1,2,\cdots,T\}$, we can conclude that 
\begin{align}
\mathrm{Tr}(\mathbf C) \geq \max_{t=1,\cdots,T} \frac{\eta(t)}{\mathrm{Tr}\left(\mathbf{Q}(t) \mathbf{v}^{}_{m_0(t)} \mathbf{v}_{m_0(t)}^{H} \mathbf{Q}^{H}(t)\right)}.
\end{align}

\subsection{The Proof of {\bf Property \ref{property:1}}}
\label{subsec:6}
According to the problem formulation \eqref{CMO}, its feasible set, denoted by $\mathcal{F}_t$, can be formulated as 
\begin{align}
     \mathcal{F}_t = \mathcal{S}_0 \cap  \mathcal{S}^1_t  \cap\mathcal{S}^2_t, \label{property1_1}
\end{align}
where 
\begin{align}
    \mathcal{S}_0 = &\{\mathbf{C}| \mathrm{Tr}({\mathbf C} ) \leq b\}\cap \{\mathbf{C}|\mathbf{C} \in \mathbb{S}_+ \},\\
    \mathcal{S}^1_t = &\bigcap_{i=1}^t \bigcap_{m=1}^M\Big \{\mathbf{C}| \mathbf v_m^H \mathbf Q(i)^H {\mathbf C} \mathbf Q(i) \mathbf v_m \nonumber\\ &~~~~~~~~~~~\leq \mathbf v_{m_0(i)}^H \mathbf Q(i)^H {\mathbf C} \mathbf Q(i) \mathbf v_{m_0(i)}\Big \},\\
     \mathcal{S}^2_t = &\bigcap_{i=1}^t\{\mathbf{C}| \eta(i) = \mathbf v_{m_0(i)}^H \mathbf Q(i)^H {\mathbf C} \mathbf Q(i) \mathbf v_{m_0(i)}\}.
\end{align}
From the formulation \eqref{property1_1}, it holds for any $t_1 > t_2$ that
\begin{align}
    &\mathcal{S}^1_{t_1} \subseteq  \mathcal{S}^1_{t_2}; \\
    &\mathcal{S}^2_{t_1} \subseteq  \mathcal{S}^2_{t_2}.
\end{align}
Therefore, it can be concluded  
\begin{align}
      \mathcal{F}_T \subseteq \mathcal{F}_{T-1} \subseteq \cdots \subseteq \mathcal{F}_{1}.
\end{align}
In addition, according to the feedback scheme aforementioned, it is easy to check that $\mathbf{C}^* \in \mathcal{S}_0, \mathbf{C}^* \in \mathcal{S}^1_t$, and  $\mathbf{C}^* \in \mathcal{S}^2_t$ hold for any $t$, which means $\mathbf{C}^* \in \mathcal{F}_t$.

\subsection{Numerical Results of Trying Different Strategies in Table \ref{table:WMD}} 
\label{subsec:7}

The BS is with antenna number $N_A=32$ and port number $N_P=8$ while the antenna number of the UE is $N_U=2$. The ground-truth channel covariance matrices are sampled from QuaDRiGa with the speed of the UE is $3$ km/h. The center frequency of the downlink channel are set as $1.275 GHZ$. In addition, Type-I codebook is built according to 5G NR standards \cite{code1}. In these experiments, the RMSE and beam precision are all chosen as the performance measure.

{Fig. \ref{fig_stragey_1}} shows the comparison between the performances of three different $m^{\prime} (t+1)$ strategies, given $\sigma_n = 1, n = 1,\cdots,N_P$ and $\mathbf{X}$ generated from the corresponding designed strategy. In {Fig. \ref{fig_stragey_1}}, The adoptions of these three strategies have similar performances in terms of both RMSE and beam precision, while the adoption of mixture strategy can outperform others in most of cases. Especially, after the $10$-th communication round, mixture strategy is the best among another the three strategies in terms of RMSE. In terms of beam precision, the same result can be obtained, that is, the mixture strategy can surpass others after the $9$-th communication round. Hence, it is a good choice that $m^{\prime} (t+1)$ is acquired by adopting mixture strategy.

\begin{figure}[!tbp]
	\centering
	\includegraphics[width= 3.5 in]{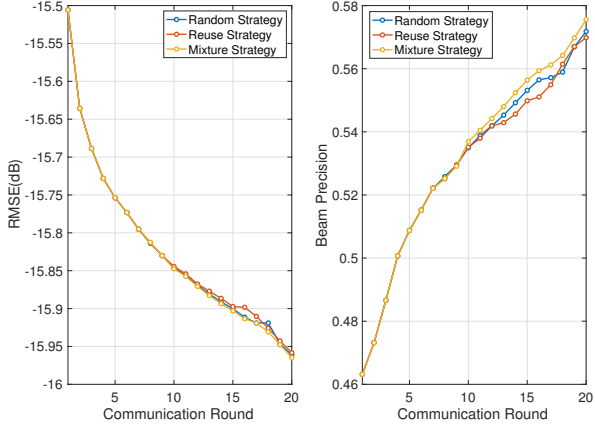}
	\caption{The RMSEs and beam precisions versus communication rounds from different strategies of choosing $m^{\prime}$, given $\sigma_n = 1, n = 1,\cdots,N_P$ and $\mathbf{X}$ generated from the corresponding designed strategy.}
	\label{fig_stragey_1}
\end{figure}

Then, given $m^{\prime}$ chosen from $[1,2,\cdots,M]$, according to the corresponding mixture strategy and $\mathbf{X}$ generated from the corresponding designed strategy, the performance of different strategies of setting $\sigma_n = 1, n = 1,\cdots,N_P$ are examined in {Fig. \ref{fig_stragey_2}} . It can be observed that the adoption of equality strategy surpasses that of the sampling-sorting strategy in terms of both RMSE and beam precision. Therefore, the best setting of $\{\sigma_n \}_{n=1}^{N_P}$ is $\sigma_1 = \sigma_2 = \cdots = \sigma_{N_P} = 1$, i.e., the equality strategy.
  
\begin{figure}[!tbp]
	\centering
	\includegraphics[width= 3.5 in]{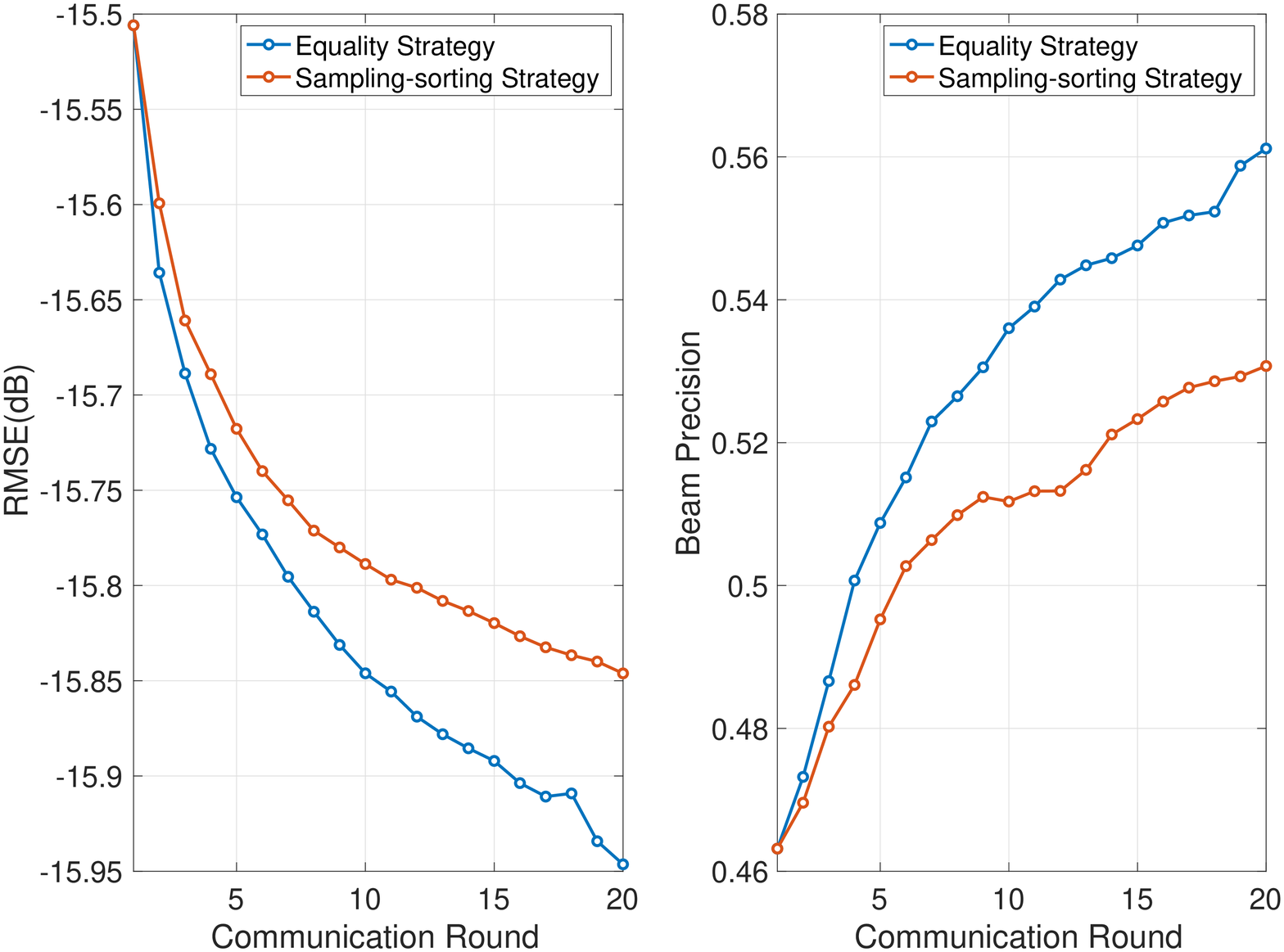}
	\caption{The RMSEs and beam precisions versus communication rounds from different strategies of choosing $\sigma_n$, given $m^{\prime}$ chose from the corresponding mixture strategy and $\mathbf{X}$ generated from the corresponding designed strategy.}
	\label{fig_stragey_2}
\end{figure}

\begin{figure}[!tbp]
	\centering
	\includegraphics[width= 3.5 in]{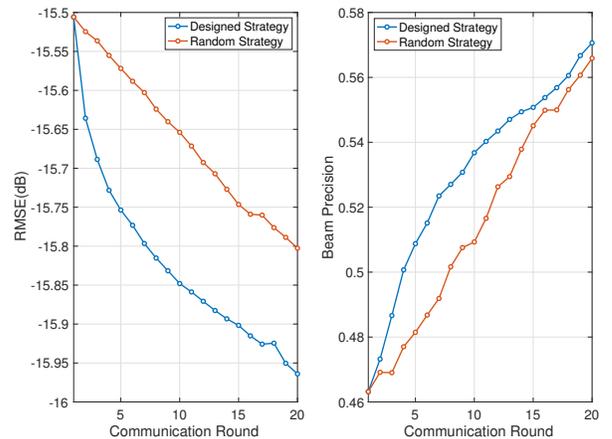}
	\caption{The RMSEs and beam precisions versus communication rounds from different strategies of choosing $\mathbf{X}$, given $m^{\prime}$ chosen from the corresponding mixture strategy and $\sigma_n = 1, n = 1,\cdots,N_P$.}
	\label{fig_stragey_3}
\end{figure} 

Finally, given $m^{\prime}$ chosen from the corresponding mixture strategy and $\sigma_n = 1, n = 1,\cdots,N_P$ selected to all $1$, the different strategies of designing $\mathbf{X}$ is compared in {Fig. \ref{fig_stragey_3}}. In terms of both RMSE or beam precision, the adoption of designed strategy surpasses another strategies. Hence, we prefer to acquire $\mathbf{X}$ according to the designed strategy.

In conclusion, the adoption of ``checked'' strategies listed in {Table \ref{table:WMD}} is the best choice in terms of both reconstruction accuracy and convergence speed.}
 




\vfill

\end{document}